\documentclass[a4paper]{IEEEtran}

\usepackage[utf8]{inputenc}

\usepackage{amsmath,amssymb,cite,dsfont}
\usepackage{mathtools}	
\usepackage{stmaryrd}	
\usepackage{bbm}	

\usepackage[usenames,dvipsnames]{xcolor} 
\usepackage{tikz}	

\usepackage{pgfplots}
\pgfplotsset{compat=1.13}

\ifCLASSOPTIONcompsoc
\usepackage[caption=false,font=normalsize,labelfont=sf,textfont=sf]{subfig}
\else
\usepackage[caption=false,font=footnotesize]{subfig}
\fi

\usepackage{hyperref}
\hypersetup{
    colorlinks,%
    citecolor=black,%
    filecolor=black,%
    linkcolor=black,%
    urlcolor=black,%
    pdfauthor={German Bassi},%
    pdftitle={Key Agreement}%
}
\usepackage[all]{hypcap}

\newcommand{\PR}[1]{\ensuremath{\textnormal{Pr}\!\left\{{#1}\right\}}}

\newcommand{\I}[2]{\ensuremath{I(#1;#2)}}
\newcommand{\IC}[3]{\ensuremath{I(#1;#2\vert #3)}}

\renewcommand{\H}[1]{\ensuremath{H(#1)}}
\newcommand{\HC}[2]{\ensuremath{H(#1\vert #2)}}
\newcommand{\mkv}{-\!\!\!\!\minuso\!\!\!\!-}
\newcommand{\cond}{\,\vert\,}

\newcommand{\cA}{\ensuremath{\mathcal A}}
\newcommand{\cB}{\ensuremath{\mathcal B}}
\newcommand{\cC}{\ensuremath{\mathcal C}}
\newcommand{\cE}{\ensuremath{\mathcal E}}

\newcommand{\cK}{\ensuremath{\mathcal K}}

\newcommand{\cQ}{\ensuremath{\mathcal Q}}
\newcommand{\cP}{\ensuremath{\mathcal P}}
\newcommand{\cR}{\ensuremath{\mathcal R}}
\newcommand{\cS}{\ensuremath{\mathcal S}}
\newcommand{\cT}{\ensuremath{\mathcal T}}
\newcommand{\cU}{\ensuremath{\mathcal U}}
\newcommand{\cV}{\ensuremath{\mathcal V}}
\newcommand{\cX}{\ensuremath{\mathcal X}}
\newcommand{\cY}{\ensuremath{\mathcal Y}}
\newcommand{\cZ}{\ensuremath{\mathcal Z}}

\newcommand{\bR}{\ensuremath{\mathbb R}}

\newcommand{\bE}{\ensuremath{\mathbb E}}

\newcommand{\lessnoisy}[1]{\succeq_{\scriptscriptstyle #1}\!}

\newcommand{\typ}[2]{\cT_\delta^{#1}(#2)}
\newcommand{\typc}[2]{\cT_{\delta'}^{#1}(#2)}

\newcommand{\ind}[1]{\ensuremath{\mathds{1} {\left\{ #1 \right\}}}}

\newtheorem{definition}	{Definition}
\newtheorem{theorem}	{Theorem}
\newtheorem{proposition}{Proposition}
\newtheorem{lemma}	{Lemma}

\newtheorem{remark}	{Remark}

\title{Secret Key Generation over Noisy Channels\\ with Correlated Sources}
\author{
  \IEEEauthorblockN{Germ{\'a}n~Bassi,~\IEEEmembership{Member,~IEEE}, Pablo~Piantanida,~\IEEEmembership{Senior Member,~IEEE}, and \\ Shlomo~Shamai~(Shitz),~\IEEEmembership{Fellow Member,~IEEE} }\\
\thanks{The work of G.~Bassi was funded in part by the Knut and Alice Wallenberg foundation and the Swedish Foundation for Strategic Research,  
  and the work of S.~Shamai was supported by the European Union's Horizon 2020 Research And Innovation Programme, grant agreement no.\ 694630.
  The material in this paper was presented in part at the 2016 IEEE International Symposium on Information Theory, Jul.\ 2016.~\cite{bassi_isit_2016}.}
\thanks{G.~Bassi is with the School of Electrical Engineering and Computer Science, KTH Royal Institute of Technology, Stockholm 100 44, Sweden (e-mail: germanb@kth.se).}
\thanks{P.~Piantanida is with CentraleSup{\'e}lec--French National Center for Scientific Research (CNRS)--Universit{\'e} Paris-Sud, 3 Rue Joliot-Curie, F-91192 Gif-sur-Yvette, France, and with Montreal Institute for Learning Algorithms (MILA) at Universit{\'e} de Montr{\'e}al, 2920 Chemin de la Tour, Montr{\'e}al, QC H3T 1N8, Canada (e-mail: pablo.piantanida@centralesupelec.fr).}
\thanks{S.~Shamai~(Shitz) is with the Department of Electrical Engineering, Technion--Israel Institute of Technology, Haifa, 32000, Israel (e-mail: sshlomo @ee.technion.ac.il).}
}

\begin{document}

\maketitle

\begin{abstract}
This paper investigates the problem of secret key generation over a wiretap channel when the terminals observe correlated sources. These sources are independent of the main channel and the users overhear them before the transmission takes place.
A novel achievable scheme is proposed, and its optimality is shown under certain less-noisy conditions. This result improves upon the existing literature where the more stringent condition of degradedness is required.
Furthermore, numerical evaluation of the proposed scheme and previously reported results for a binary model are presented; a comparison of the numerical bounds provides insights on the benefit of the novel scheme.
\end{abstract}


\section{Introduction}

\IEEEPARstart{T}{he wiretap} channel, introduced by Wyner~\cite{wyner_wiretap_1975}, is the basic model for analyzing secrecy in wireless communications. In this model, the transmitter, named Alice, wants to communicate reliably with Bob while keeping the transmitted message --or part of it-- secret from an eavesdropper, named Eve, overhearing the communication through another channel. Secrecy is characterized by the amount of information that is not \emph{leaked}, which can be measured by the equivocation rate --the remaining uncertainty about the message at the eavesdropper. The secrecy capacity of the wiretap channel is thus defined as the maximum transmission rate that can be attained with zero leakage. In their influential paper~\cite{csiszar_broadcast_1978}, Csisz{\'a}r and K{\"o}rner determine the rate-equivocation region of a general broadcast channel with any arbitrary level of security, which also establishes the secrecy capacity of the wiretap channel. These schemes guarantee secrecy by exploiting an artificial random noise that saturates the eavesdropper's decoding capabilities. 

On the other hand, Shannon~\cite{shannon_secrecy_1949} shows that it is also possible to achieve a positive secrecy rate by means of a \emph{secret key}. Alice and Bob can safely communicate over a noiseless public broadcast channel as long as they share a secret key. The rate of this key, however, must be at least as large as the rate of the message to attain zero leakage. The main question that arises in this scenario is therefore: how do the legitimate users safely share the secret key? The answer is that the users should not communicate the key itself, which would then be compromised. Instead, they should only convey enough information to allow themselves to \emph{agree} upon a key without disclosing, at the same time, any relevant information about it to the eavesdropper (for further discussion the reader is referred to~\cite{chorti_physical_2016, narayan_tutorial_2016}).

\begin{figure}[t!]
\centering
\begin{tikzpicture}[line width=1pt, font=\footnotesize, scale=1]
\draw[latex-,thick] (0.35,2.6) -- (1,2.6) node[midway,above] {$\,K$};
\draw (1,2.2) rectangle (2,3) node[pos=.5] {Alice};
\draw[-latex,thick] (2,2.6) -- (2.8,2.6) node[midway,above] {$X^m$};
\draw (2.8,2.2) rectangle (4.6,3) node[pos=.5] {$p(yz\vert x)$};
\draw[-latex,thick] (4.6,2.6) -- (5.4,2.6) node[midway,above] {$Y^m$};
\draw[-latex,thick] (3.7,2.2) -- (3.7,1.55) -- (4.7,1.55) node[midway,above] {$Z^m$};
\draw (5.4,2.2) rectangle (6.4,3) node[pos=.5] {Bob};
\draw[-latex,thick] (6.4,2.6) -- (6.95,2.6) node[midway,above] {$\hat{K}$} -- (7,2.6) node[right] {$\PR{\hat{K}\!\neq\!K} \!\leq\! \epsilon$};
\draw (4.7,1.15) rectangle (5.7,1.95) node[pos=.5] {Eve};
\draw[-latex,thick] (5.7,1.55) -- (6.3,1.55) node[right] {$\I{K}{Z^m E^n} \leq n\epsilon$};
\draw[Blue] (1.9,0.5) rectangle (3.5,1.3) node[pos=.5] {$p(abe)$};
\draw[-latex,thick,Blue] (1.9,.9) -- (1.5,.9) -- (1.5,2.2) node[left,midway] {$A^n$};
\draw[-latex,thick,Blue] (3.5,.85) -- (3.7,.85) -- (4.7,.85) node[above,midway] {$E^n$} -- (5.2,.85) -- (5.2,1.15);
\draw[-latex,thick,Blue] (2.7,.5) -- (2.7,.35) -- (5.9,.35) -- (5.9,1.45) node[right,midway] {$B^n$} arc (-90:90:.1) -- (5.9,2.2);
\end{tikzpicture}
\caption{System model for the problem of secret key generation.
\label{fig:model_key}}
\end{figure}
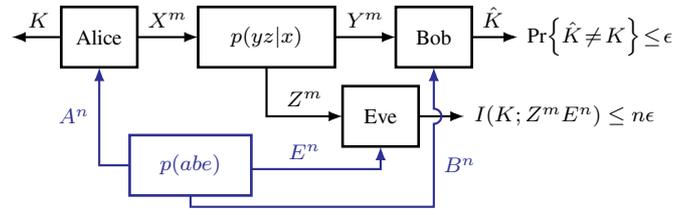

In this work, we study the problem of secret key generation over a wiretap channel with correlated sources at each terminal.
These sources are assumed to be independent of the main channel and there is no additional public broadcast channel of finite or infinite rate, as seen in Fig.~\ref{fig:model_key}.
It is assumed that each node acquires the $n$-sequence observation of its corresponding source before the communication begins.

\subsection{Related Work}

Maurer~\cite{maurer_secret_1993} and Ahlswede and Csisz{\'a}r~\cite{ahlswede_common_1993} are among the first to study the use of correlated observations available at the legitimate users as a means to agree upon a key. In addition to the correlated observations, the terminals may communicate over a public broadcast channel of infinite capacity to which the eavesdropper has also access. Two models are proposed in~\cite{ahlswede_common_1993}: the ``source model'', where the users observe correlated random sources controlled by nature, and the ``channel model'', where the users observe inputs and outputs of a noisy channel controlled by one of the users.
In~\cite{csiszar_common_2000}, Csisz{\'a}r and Narayan study the first model but assume that the public broadcast channel has finite capacity and there is a third ``helper'' node who is not interested in recovering the key but rather helping Alice and Bob. The same authors also analyze the channel model with only one~\cite{csiszar_secrecy_2008} or with multiple channel inputs~\cite{csiszar_secrecy_2013}. Capacity results are presented in~\cite{csiszar_common_2000, csiszar_secrecy_2008, csiszar_secrecy_2013} assuming that there is only one round of communication over the public channel. General inner and outer bounds for both source and channel models with interaction over the public channel are introduced by Gohari and Anantharam in~\cite{gohari_sk-source_2010, gohari_sk-channel_2010}.

More recently, Khisti \emph{et al.}~\cite{khisti_secret-key_2012} investigate the situation where there is no helper node, the users communicate over a wiretap channel, and a separate public discussion channel may or may not be available. The simultaneous transmission of a secret message along with a key generation scheme using correlated sources is analyzed by Prabhakaran \emph{et al.}~\cite{prabhakaran_secrecy_2012}. The authors obtain a simple expression that reveals the trade-off between the achievable secrecy rate and the achievable rate of the secret key. 
The corresponding Gaussian channel with correlated Gaussian sources but independent of the channel components is recently studied in~\cite{bunin_gaussian_2016}. Closed form expressions for both secret key generation and secret message transmission are derived. 
On the other hand, Salimi \emph{et al.}~\cite{salimi_key_2013} consider simultaneous key generation of two independent users over a multiple access channel with feedback, where each user eavesdrops the other. In addition, the receiver can actively send feedback, through a private noiseless (or noisy) link, to increase the size of the shared keys.

The authors of~\cite{khisti_secret-key_2012, prabhakaran_secrecy_2012, bunin_gaussian_2016} do not assume interactive communication, i.e., there is only one round of communication. Salimi \emph{et al.}~\cite{salimi_key_2013}, however, allow the end user to respond once through the feedback link.
Other authors have analyzed key generation schemes that rely on several rounds of transmissions.
Tyagi~\cite{tyagi_common_2013} characterizes the minimum communication rate required to generate a maximum-rate secret key with $r$ rounds of interactive communication. He shows that this rate is equal to the \emph{interactive common information} (a quantity he introduces) minus the secret key capacity. In his model, two users observe i.i.d. correlated sources and communicate over an error-free channel. 
Hayashi \emph{et al.}~\cite{hayashi_secret_2016} study a similar problem but consider general (not necessarily i.i.d.) source sequences of finite length. Their proposed protocol attains the secret key capacity for general observations as well as the second-order asymptotic term of the maximum feasible secret key length for i.i.d. observations. They also prove that the standard one-way communication protocol fails to attain the aforementioned asymptotic result.
Courtade and Halford~\cite{courtade_coded_2016} analyze the related problem of how many rounds of public transmissions are required to generate a specific number of secret keys. Their model assumes that there are $n$ terminals connected through an error-free public channel, where each terminal is provided with a number of messages before transmission that it uses to generate the keys.

As previously mentioned, the focus of the present work is on sources that are independent of the main channel; nonetheless, some works have addressed the general situation of correlated sources and channels. Prior work on secrecy for channels with state include Chen and Vinck's~\cite{chen_wtc_2008} and Liu and Chen's~\cite{liu_wiretap_2007} analysis of the wiretap channel with state. These works employ Gelfand and Pinsker's scheme~\cite{gelfand_coding_1980} to correlate the transmitted codeword with the channel state at the same time that it saturates the eavesdropper's decoding capabilities.
A single-letter expression of the secrecy capacity for this model is still unknown, although a multi-letter bound is provided by Muramatsu~\cite{muramatsu_general_2014} and a novel lower bound is recently reported in~\cite{goldfeld_noncausally_2016}. As a matter of fact, the complexity of this problem also lies in the derivation of an outer bound that can handle simultaneously secrecy and channels with state.

To the best of our knowledge, only a handful of works have studied the problem of key generation for channels with state. The previously mentioned result of Prabhakaran \emph{et al.}~\cite{prabhakaran_secrecy_2012} is one of these examples. Zibaeenejad~\cite{zibaeenejad_key_2015} analyzes a similar scenario where there is also a public channel of finite capacity between the users and he provides an inner and an outer bound for this model.
Although the inner bound is developed for a channel with state, it is possible to apply it to the model used in the present work, i.e., sources independent of the main channel. However, some steps of the proof reported in~\cite{zibaeenejad_key_2015} appear to be obscure and a constraint seems to be missing in the final expression; the resulting achievable rate was recently shown in~\cite{bunin_key_2018} to be in certain cases unachievable. As a consequence, we have decided not to compare our inner bound to this previously reported scheme.

The works found in the literature that are closely related to the problem dealt here~\cite{khisti_secret-key_2012, prabhakaran_secrecy_2012, bunin_gaussian_2016, salimi_key_2013, zibaeenejad_key_2015} derive the equivocation of their schemes using a \emph{weak secrecy} condition.
%
%
In line with these works, we use the same measure in the analysis of our proposed scheme; however, it can be shown that our strategy also fulfills the \emph{strong secrecy} criterion (see Remark~\ref{rk:strong_secrecy}) which has become more frequent nowadays.
Recent works on the wiretap channel employ this approach, e.g., \cite{wiese_strong_2016, goldfeld_semantic_2016}, where in particular~\cite{goldfeld_semantic_2016} does not assume that the messages have a uniform distribution.

Finally, it is worth noting that the problem of secure source transmission with side information~\cite{ekrem_secure_2011, villard_secure_2013, villard_secure_2014} is closely related to the present work, since the reconstructed source may serve as a key as long as it has been reliably and securely transmitted. It is not surprising that some of the techniques developed in those works may be found here as well.

\subsection{Contributions and Organization of the Paper}

In this work, we introduce a novel coding scheme (Theorem~\ref{th:digital-key}) for the problem of secret key generation over a wiretap channel with correlated sources at each terminal. The correlated sources are assumed to be independent of the main channel and, thanks to a previously reported outer bound~\cite{bassi_wiretap_2015}, this scheme is shown to be optimal (Propositions~\ref{prop:sp_ln_z}, \ref{prop:sp_ln_e}, and~\ref{prop:sp_ln_yb}) whenever the channel and/or source components satisfy the specific \emph{less-noisy} conditions described in Table~\ref{tab:optimal}.
In contrast, the proposed schemes in~\cite{khisti_secret-key_2012, prabhakaran_secrecy_2012, bunin_gaussian_2016, salimi_key_2013} were optimal only when the stronger \emph{degradedness} condition holds true for the channel and source components.

The main improvement of our scheme with respect to the literature is to introduce a two-layer codebook for describing the source. Although a two-layer scheme is not a new technique for the ``source model'' (cf.~\cite[Thm.~1]{ahlswede_common_1993}), it introduces considerable difficulty and has not been investigated in the framework of the combined model of Fig.~\ref{fig:model_key}. Difficulty arises in the derivation of Eve's equivocation, as shown by Lemma~\ref{lem:lem_fano_eve} in Section~\ref{ssec:key_leakage}.
However, a scheme that is developed with two description layers can achieve higher secret key rates than those of a single-layer scheme.

This paper is organized as follows. Section~\ref{sec:prelim} provides some definitions and our previously reported outer bound. In Section~\ref{sec:main_result}, we first present the inner bound for the problem of secret key agreement and then we enumerate the cases where said achievable scheme is optimal. Section~\ref{sec:bec_bsc_channel} illustrates with a binary example the improvement of the present work over a previously reported scheme. In Section~\ref{sec:proof-digital}, we give the detailed proof of the inner bound. Finally, Section~\ref{sec:summary} summarizes and concludes the work.

\subsection*{Notation and Conventions}

Throughout this work, we use the standard notation of~\cite{gamal_network_2011}.
Specifically, given two integers $i$ and $j$, the expression $[i: j]$ denotes
the set $\{i, i+1, \ldots, j\}$, whereas for real values $a$ and $b$, $[a, b]$
denotes the closed interval between $a$ and $b$.
We use the notation $x_i^j = (x_i, x_{i+1}, \ldots, x_j)$ to denote the
sequence of length $j-i+1$ for $1\leq i\leq j$. If $i=1$, we drop the subscript
for succinctness, i.e., $x^j = (x_1, x_2, \ldots, x_j)$.
Lowercase letters such as $x$ and $y$ are mainly used to represent constants or realizations of random variables,
capital letters such as $X$ and $Y$ stand for the random variables in itself,
and calligraphic letters such as $\cX$ and $\cY$ are reserved for sets, codebooks, or special functions.

The set of nonnegative real numbers is denoted by $\bR_+$. 
The probability distribution~(PD) of the random vector $X^n$, $p_{X^n}(x^n)$, is
succinctly written as $p(x^n)$ without subscript when it can be understood from
the argument $x^n$.
Given three random variables $X$, $Y$, and $Z$, if its joint PD can be
decomposed as $p(xyz) = p(x) p(y\vert x) p(z\vert y)$, then they form a Markov
chain, denoted by $X \mkv Y \mkv Z$.
The random variable $Y$ is said to be \emph{less noisy} than $Z$ w.r.t.\ $X$ if $\I{U}{Y}\geq \I{U}{Z}$ for each random variable $U$ such that $U\mkv X\mkv (Y,Z)$; this relation is denoted by $Y\lessnoisy{X}Z$.
Entropy is denoted by $\H{\cdot}$ and mutual information, $\I{\cdot}{\cdot}$. The expression $[x]^+$ denotes $\max\{ x, 0 \}$.
Given $u,v\in[0,1]$, the function $h_2(u) \triangleq -u\log_2 u -(1-u)\log_2(1-u)$ is the binary entropy function and $u*v \triangleq u(1-v)+v(1-u)$.
We denote typical and conditional typical sets by $\typ{n}{X}$ and $\typ{n}{Y|x^n}$, respectively.

\section{Preliminaries}
\label{sec:prelim}

\subsection{Problem Definition}

Consider the wiretap channel with correlated sources at every node $(A,B,E)$, as shown in Fig.~\ref{fig:model_key}. The legitimate users (Alice and Bob) want to agree upon a secret key $K \in \cK$ while an eavesdropper (Eve) is overhearing the communication.
Let $\cA$, $\cB$, $\cE$, $\cX$, $\cY$, and $\cZ$ be six finite sets. Alice, Bob, and Eve observe the random sequences (sources) $A^n$, $B^n$, and $E^n$, respectively, drawn i.i.d. according to the joint distribution $p(abe)$ on $\cA\times\cB\times\cE$.
Alice communicates with Bob through $m$ instances of a discrete memoryless channel with input $X\in\cX$ and output $Y\in\cY$. Eve is listening the communication through another channel with input $X\in\cX$ and output $Z\in\cZ$. This channel is defined by its transition probability $p(yz\vert x)$ and it is independent of the sources' distribution.


\vspace*{1mm}
\begin{definition}[Code]
\label{def:code}
A $(2^{nR_k},n,m)$ secret key code $\mathsf{c}_n$ for this model consists of:
\begin{itemize}
 \item a key set $\cK_n\triangleq[1:2^{nR_k}]$,
 \item a source of local randomness $R_r\in\cR_r$ at Alice,
 \item an encoding function $\varphi\colon \cA^n\times\cR_r \to \cX^m$,
 \item a key generation function $\psi_a\colon \cA^n\times\cR_r \to \cK_n$, and
 \item a key generation function $\psi_b\colon \cB^n\times\cY^m \to \cK_n$.
\end{itemize}
The rate of such a code is defined as the number of channel uses per source symbol~$\frac{m}{n}$.
\end{definition}
\vspace{1mm}

Given a code, let $K=\psi_a(A^n, R_r)$ and $X^m=\varphi(A^n, R_r)$; then, the performance of the $(2^{nR_k},n,m)$ secret key code $\mathsf{c}_n$ is measured in terms of its average probability of error
\begin{equation}
\mathsf{P}_{\!e}(\mathsf{c}_n) \triangleq \PR{ \psi_b(B^n,Y^m) \neq K \vert \mathsf{c}_n},
\label{eq:def_sk_rel}
\end{equation}
in terms of the information leakage
\begin{equation}
\mathsf{L}_k(\mathsf{c}_n) \triangleq \IC{K}{E^n Z^m}{\mathsf{c}_n}\,,
\label{eq:def_sk_leak}
\end{equation}
and in terms of the uniformity of the keys
\begin{equation}
\mathsf{U}_k(\mathsf{c}_n) \triangleq nR_k -\HC{K}{\mathsf{c}_n}\,.
\label{eq:def_sk_unif}
\end{equation}

\vspace{1mm}
\begin{definition}[Achievability]
\label{def:achievability}
A tuple $(\eta,R_k)\in\bR_+^2$ is said to be \emph{achievable} for this model if, for every $\epsilon>0$ and sufficiently large $n$, there exists a $(2^{nR_k},n,m)$ secret key code $\mathsf{c}_n$ such that
\begin{equation}
 \frac{m}{n}                    \leq \eta+\epsilon\,,\ \
 \mathsf{P}_{\!e}(\mathsf{c}_n) \leq \epsilon\,,\ \
 \frac{1}{n}\mathsf{L}(\mathsf{c}_n) \leq \epsilon\,,\ \
 \frac{1}{n}\mathsf{U}(\mathsf{c}_n) \leq \epsilon\,.
 \label{eq:sk_rate_weak}
\end{equation}

The set of all achievable tuples is denoted by $\cR^\star$ and is referred to as the \emph{secret key region}.%
\end{definition}

\subsection{Outer Bound}

The following theorem gives an outer bound on $\cR^\star$, i.e., it defines the region $\cR_\text{out}\supseteq\cR^\star$.

\vspace*{1mm}
\begin{theorem}
\label{th:outer}
An \emph{outer bound} on the secret key region for this channel model is given by
\begin{multline}
R_k \leq \max_{p\in\cP} \big\{ \eta \big[ \I{T}{Y} -\I{T}{Z} \big] \\+\IC{V}{B}{U} -\IC{V}{E}{U} \big\} \label{eq:outer-rate}
\end{multline}
subject to
\begin{equation}
\IC{V}{A}{B} \leq \eta\,\I{X}{Y}\,, \label{eq:outer-cond}
\end{equation}
where $\cP$ is the set of input probability distributions given by%
\begin{multline}
\cP = \big\{\, p( tx yz uv abe ) = \\p(t x) p(yz \vert x) p(abe) p( v \vert a ) p( u \vert v ) \,\big\} \label{eq:outer-pmf}
\end{multline}
with $|\cT| \leq |\cX|$, $|\cU| \leq |\cA| +1$, and $|\cV| \leq (|\cA| +1)^2$.
\end{theorem}
\begin{IEEEproof}
Refer to Appendix~\ref{sec:Proof-OB} for details.
\end{IEEEproof}
\vspace*{1mm}

Theorem~\ref{th:outer} shows that the secret key generated between Alice and Bob has two components. 
The first two terms on the r.h.s. of~\eqref{eq:outer-rate} represent the part of the key that is securely transmitted through the noisy channel (given by the random variable $T$) as in the wiretap channel. On the other hand, the last two terms on the r.h.s. of~\eqref{eq:outer-rate} characterize the part of the key that is securely extracted from the correlated sources (given by the random variables $U$ and $V$).
Since the source and channel variables are independent in the model, it should not be surprising that the variable $T$ is independent of $(U,V)$. However, given that the users need to agree on common extracted bits from the source, the noisy channel imposes the restriction~\eqref{eq:outer-cond} on the amount of information exchanged during that agreement.

\vspace*{1mm}
\begin{remark}
The calculation of the bounds~\eqref{eq:outer-rate} and~\eqref{eq:outer-cond} is done using the probability distribution~\eqref{eq:outer-pmf}.
However, we note that~\eqref{eq:outer-pmf} is an uncommon single-letter expression of the source and channel variables since the sequences have different lengths. This remark is also applicable to all the regions presented in the sequel.
\end{remark}

\section{Main Results}
\label{sec:main_result}

In this section, we first introduce a key generation scheme for the aforementioned model that leads to a novel inner bound for the secret key region (Theorem~\ref{th:digital-key}). We then study some special cases where this scheme turns out to achieve the (optimal) secret key region (Propositions~\ref{prop:sp_ln_z}, \ref{prop:sp_ln_e}, and~\ref{prop:sp_ln_yb}).

\subsection{Inner Bound}
\label{ssec:inner_bound}

The following theorem gives an inner bound on $\cR^\star$, i.e., it defines the region $\cR_\text{in}\subseteq\cR^\star$.

\vspace*{1mm}
\begin{theorem}\label{th:digital-key}
A tuple $(\eta,R_k)\in\bR_+^2$ is achievable if there exist random variables $U$, $V$, $Q$, $T$, $X$ on finite sets \cU, \cV, \cQ, \cT, \cX, respectively, with joint distribution $p(uv qtx yz abe) = p(q\vert t)p(tx)p(yz\vert x) p(abe)p(v\vert a)p(u\vert v)$, which verify 
\begin{multline}
R_k \leq \eta\big[ \IC{T}{Y}{Q} -\IC{T}{Z}{Q} \big] \\ +\IC{V}{B}{U} -\IC{V}{E}{U} \label{eq:th1-main_result_rate}
\end{multline}
subject to
\begin{subequations}\label{eq:th1-main_result}
\begin{align}
 \IC{U}{A}{B} &\leq \eta\,\I{Q}{Y}\,, \label{eq:th1-main_result1}\\
 \IC{V}{A}{B} &\leq \eta\,\I{T}{Y}\,. \label{eq:th1-main_result2}
\end{align}
\end{subequations}
Moreover, it suffices to consider sets $\cU$, $\cV$, $\cQ$, and $\cT$ such that $|\cU| \leq |\cA|+2$, $|\cV| \leq (|\cA|+1)(|\cA|+2)$, $|\cQ| \leq |\cX|+2$, and $|\cT| \leq (|\cX|+1)(|\cX|+2)$.
\end{theorem}
\begin{IEEEproof}
Alice employs the two-layer description $(U,V)$ to compress the source $A$ and it transmits it through the two-layer channel codeword $(Q,T)$. Each layer of the description must fit in the corresponding layer of the channel codeword according to~\eqref{eq:th1-main_result}. The achievable secret key rate~\eqref{eq:th1-main_result_rate} is a combination of the secret bits transmitted through the noisy channel in the manner of the wiretap channel and the secret bits obtained by the reconstruction of the source at Bob.
The full proof is deferred to Section~\ref{sec:proof-digital}.
\end{IEEEproof}

\vspace{1mm}
\begin{remark}
The regions $\cR_\text{out}$ and $\cR_\text{in}$ do not coincide in general. This is due to the presence of the condition~\eqref{eq:th1-main_result1} in the inner bound, and the looser condition~\eqref{eq:outer-cond} in the outer bound with respect to~\eqref{eq:th1-main_result2}.
We present in Section~\ref{ssec:optimal} a few special cases where these differences disappear and both regions coincide. 
\end{remark}

\vspace{1mm}
\begin{remark}
\label{rk:inner_sep_prev_results}
By setting $U=\emptyset$, the region in Theorem~\ref{th:digital-key} recovers the results in~\cite[Thm.~1 and~4]{khisti_secret-key_2012}, when the eavesdropper has access to a correlated source, and \cite[Thm.~2]{prabhakaran_secrecy_2012}, when there is no secret message to be transmitted. In these works, there was only one layer to encode the source $A^n$ while our coding scheme allows for two layers, introducing considerable difficulty in the derivation of Eve's equivocation (see e.g. the multiple binning stages in the proof and~\eqref{eq:th1-keyleak_6}).
The advantage of having two layers of description is that Theorem~\ref{th:digital-key} can potentially achieve higher secret key rates (see Section~\ref{sec:bec_bsc_channel}) and it recovers the result of Csisz{\'a}r and Narayan~\cite{csiszar_common_2000} (see Remark~\ref{rk:csiszar_common}).
\end{remark}

\vspace{1mm}
\begin{remark}
The region in Theorem~\ref{th:digital-key} also recovers the result in~\cite[Thm.~1]{cao_secret_2017}, which was published after the original submission of this manuscript. In that work, Alice and Bob communicate over a public noiseless channel of rate $R_1$ and a secure noiseless channel of rate $R_2$. The proposed achievable scheme in~\cite{cao_secret_2017} sends the codeword $Q$ through the public channel, i.e., $\I{Q}{Y}=R_1$, and the codeword $T$ through the secure channel, i.e., $\IC{T}{Y}{Q}=R_2$ and $\IC{T}{Z}{Q}=0$. The reader may verify that, by using the aforementioned quantities and $\eta=1$, both regions are equal.
\end{remark}

\vspace{1mm}
\begin{remark}
Theorem~\ref{th:digital-key} improves upon our previous work in~\cite[Sec.~IV-A]{bassi_wiretap_2015} since~\eqref{eq:th1-main_result} replaces the more stringent condition: $\IC{V}{A}{B} \leq \eta\,\I{Q}{Y}$.
\end{remark}

\vspace{1mm}
\begin{remark}
The problem of key generation dealt with in the present work is intimately connected to the problem of secure source transmission with side information, at both receiver and eavesdropper~\cite{villard_secure_2013, villard_secure_2014}, since the part of the source that can be reliably and securely transmitted serves as key which is a function of the source. It is thus not surprising that Theorem~\ref{th:digital-key} bears a resemblance to our previous result in~\cite[Thm.~2]{villard_secure_2014}.
\end{remark}

\vspace{1mm}
\begin{remark}
\label{rk:strong_secrecy}
Theorem~\ref{th:digital-key} is obtained using the \emph{weak} secrecy and uniformity conditions in~\eqref{eq:sk_rate_weak}.
However, employing the method introduced in~\cite{maurer_strong_2000}, we can show that the \emph{strong} secrecy and uniformity conditions, i.e., $\mathsf{L}(\mathsf{c}_n) \leq \epsilon$ and $\mathsf{U}(\mathsf{c}_n) \leq \epsilon$, also hold true.
The proof relies on using $l$ times a secret key code $\mathsf{c}_n$ to generate $l$ independent keys. We then interpret these $l$ keys as $l$ realizations of a random source in the ``source model'', which allows us to distill strong secret keys by means of a one-way direct reconciliation protocol and privacy amplification with extractors.
These two steps involve the transmission of additional information through the channel; nonetheless, the cost of these additional channel uses is negligible compared to the total transmission time for large $l$, $m$, and $n$.
We omit the details of the proof here due to the similarities with~\cite[Sec.~4.5]{bloch_physical_2011} and~\cite[App.~B-C]{bassi_wiretap_2015}.
\end{remark}

\subsection{Optimal Characterization of the Secret Key Rate}
\label{ssec:optimal}

The inner bound $\cR_\text{in}$ is optimal under certain less-noisy conditions in channel and/or source components. These special cases are summarized in Table~\ref{tab:optimal}.

\begin{table}[!t]
{%
\newcommand{\mc}[3]{\multicolumn{#1}{#2}{#3}}
\renewcommand{\arraystretch}{1.3} 
\begin{center}
\begin{tabular}{|c|c|c|}
\cline{2-3}
\mc{1}{c|}{}      & $E\lessnoisy{A}B$              & $B\lessnoisy{A}E$\\
\hline
$Z\lessnoisy{X}Y$ & $R_k=0$                        & Proposition~\ref{prop:sp_ln_z}\\
\hline
$Y\lessnoisy{X}Z$ & Proposition~\ref{prop:sp_ln_e} & Proposition~\ref{prop:sp_ln_yb}\\
\hline
\end{tabular}
\end{center}
}%
\caption{Regimes where Theorem~\ref{th:digital-key} is optimal. No secret key is achievable if $Z\lessnoisy{X}Y$ and $E\lessnoisy{A}B$.
\label{tab:optimal}}
\end{table}

\vspace*{1mm}
\subsubsection{Eve Has a Less Noisy Channel}

If Eve has a less noisy channel than Bob, i.e., $Z\lessnoisy{X}Y$, the information transmitted over the channel is compromised. Therefore, the amount of secret key that can be generated only depends on the statistical differences between sources. 

\vspace*{1mm}
\begin{proposition}\label{prop:sp_ln_z}
If $Z\lessnoisy{X}Y$, a tuple $(\eta,R_k)\in\bR_+^2$ is achievable if and only if there exist random variables $U$, $V$, $X$ on finite sets \cU, \cV, \cX, respectively, with joint distribution $p(uvabexyz) = p(u\vert v)p(v\vert a)p(abe) p(x)p(yz\vert x)$, which verify 
\begin{subequations}\label{eq:sp_ln_z}
\begin{gather}
 R_k \leq \IC{V}{B}{U} -\IC{V}{E}{U} \\
 \textnormal{subject to }\ \IC{V}{A}{B} \leq \eta\,\I{X}{Y}\,.
\end{gather}
\end{subequations}
\end{proposition}
\begin{IEEEproof}
Given the less-noisy condition on Eve's channel, i.e., $\I{T}{Y} \leq \I{T}{Z}$ for any RV $T$ such that $T\mkv X\mkv (YZ)$, the bound~\eqref{eq:outer-rate} is maximized with $T=\emptyset$.
On the other hand, the region~\eqref{eq:sp_ln_z} is achievable by setting auxiliary RVs $Q=T=X$ in $\cR_\text{in}$.
\end{IEEEproof}

\vspace{1mm}
\begin{remark}\label{rk:csiszar_common}
The secret key capacity of the wiretap channel with a public noiseless channel of rate $R$~\cite[Thm.~2.6]{csiszar_common_2000} turns out to be a special case of Proposition~\ref{prop:sp_ln_z}, where $X = Y = Z$ and defining $\eta\,\H{X}=\eta \log |\cX| \equiv R$.
\end{remark}


\vspace*{1mm}
\subsubsection{Eve Has a Less Noisy Source}

If Eve has a less noisy source than Bob, i.e., $E\lessnoisy{A}B$, the amount of secret key that can be generated depends on the amount of secure information transmitted through the wiretap channel.

\vspace*{1mm}
\begin{proposition}\label{prop:sp_ln_e}
If $E\lessnoisy{A}B$, a tuple $(\eta,R_k)\in\bR_+^2$ is achievable if and only if there exist random variables $T$, $X$ on finite sets \cT, \cX, respectively, with joint distribution $p(txyz) = p(tx)p(yz\vert x)$, which verify
\begin{equation}
 R_k \leq \eta\big[ \I{T}{Y} -\I{T}{Z} \big]\,. \label{eq:sp_ln_e}
\end{equation}
\end{proposition}
\begin{IEEEproof}
Given the less-noisy condition on Eve's source, i.e., $\I{V}{B} \leq \I{V}{E}$ for any RV $V$ such that $V\mkv A\mkv (BE)$, the bound~\eqref{eq:outer-rate} is maximized with $U=V$ and independent of the sources.
The region~\eqref{eq:sp_ln_e} is achievable by using the same auxiliary RVs in the inner bound as in the outer bound.
\end{IEEEproof}

\vspace{1mm}
\begin{remark}
The bound~\eqref{eq:sp_ln_e} is equal to the secrecy capacity of the wiretap channel.
\end{remark}

\vspace{1mm}
\begin{remark}
Even though the bound~\eqref{eq:sp_ln_e} becomes independent of the sources sequences $(A^n,B^n,E^n)$, we assume $n\neq 0$, and thus the rate $\eta$ is finite.
\end{remark}

\vspace*{1mm}
\subsubsection{Bob Has a Less Noisy Channel and Source}

If Bob has a less noisy channel and source than Eve, i.e., $Y\lessnoisy{X}Z$ and $B\lessnoisy{A}E$, the lower layers of the channel and source codewords are not needed any more.

\vspace*{1mm}
\begin{proposition}\label{prop:sp_ln_yb}
If $Y\lessnoisy{X}Z$ and $B\lessnoisy{A}E$, a tuple $(\eta,R_k)\in\bR_+^2$ is achievable if and only if there exist random variables $V$, $X$ on finite sets \cV, \cX, respectively, with joint distribution $p(vabexyz) = p(v\vert a)p(abe) p(x)p(yz\vert x)$, which verify 
\begin{subequations}\label{eq:sp_ln_yb}
\begin{gather}
 R_k \leq \eta\big[ \I{X}{Y} -\I{X}{Z} \big] +\I{V}{B} -\I{V}{E}\\
 \textnormal{subject to }\ \IC{V}{A}{B} \leq \eta\,\I{X}{Y}\,.
\end{gather}
\end{subequations}
\end{proposition}
\begin{IEEEproof}
Given the less-noisy conditions on Bob's channel and source, the bound~\eqref{eq:outer-rate} is maximized with $U=\emptyset$ and $T=X$.
The region~\eqref{eq:sp_ln_yb} is achievable by also setting auxiliary RVs $U=Q=\emptyset$ and $T=X$ in the inner bound.
\end{IEEEproof}

\vspace{1mm}
\begin{remark}
Proposition~\ref{prop:sp_ln_yb} extends the results from~\cite[Thm.~4]{khisti_secret-key_2012} and~\cite[Thm.~3]{prabhakaran_secrecy_2012} which assumed the more stringent conditions of degradedness: $A\mkv B\mkv E$ and $X\mkv Y\mkv Z$.
\end{remark}

\section{Secret Key Agreement over a Wiretap Channel with BEC/BSC Sources}
\label{sec:bec_bsc_channel}

As mentioned in Remark~\ref{rk:inner_sep_prev_results}, the inner bound introduced in Section~\ref{ssec:inner_bound} employs two layers of description, and thus it is an improvement over previously reported results.
In this section, we compare the performance of our achievable scheme with that of~\cite{khisti_secret-key_2012} for a specific binary source and channel model.

\subsection{System Model}

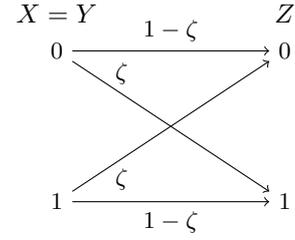
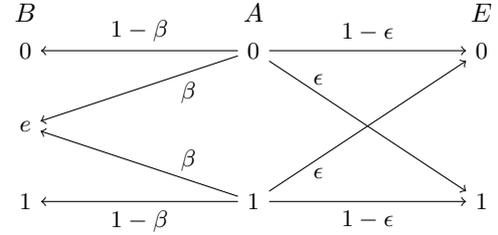
\begin{figure}[!t]
\centering
\subfloat[Main channel.]{%
\begin{tikzpicture}
	\node		(X) 	at (0,.5) 	{$X=Y$};
	\node		(X0) 	at (0,0) 	{\small $0$};
	\node		(X1)	at (0,-2)	{\small $1$};

	\node		(Z) 	at (3,.5) 	{$Z$};
	\node		(Z0) 	at (3,0) 	{\small $0$};
	\node		(Z1)	at (3,-2)	{\small $1$};
	
	\draw[->] (X0) to node[midway,above]		{\small $1-\zeta$}	(Z0);
	\draw[->] (X0) to node[near start,above]	{\small $\zeta$}	(Z1);
	\draw[->] (X1) to node[near start,below]	{\small $\zeta$}	(Z0);
	\draw[->] (X1) to node[midway,below]		{\small $1-\zeta$}	(Z1);
\end{tikzpicture}
\label{fig:bec_bsc_channel_a}
}
\hfil
\subfloat[BEC/BSC sources.]{%
\begin{tikzpicture}
	\node	(A) 	at (0,.5) 	{$A$};
	\node	(A0) 	at (0,0) 	{\small $0$};
	\node	(A1)	at (0,-2)	{\small $1$};

	\node	(B) 	at (-3,.5) 	{$B$};
	\node	(B0) 	at (-3,0) 	{\small $0$};
	\node	(Be)	at (-3,-1)	{\small $e$};
	\node	(B1)	at (-3,-2)	{\small $1$};

	\node	(E) 	at (3,.5) 	{$E$};
	\node	(E0) 	at (3,0) 	{\small $0$};
	\node	(E1)	at (3,-2)	{\small $1$};
	
	\draw[->] (A0) to node[midway,above]		{\small $1-\epsilon$}		(E0);
	\draw[->] (A0) to node[near start,above]	{\small $\epsilon$}		(E1);
	\draw[->] (A1) to node[near start,below]	{\small $\epsilon$}		(E0);
	\draw[->] (A1) to node[midway,below]		{\small $1-\epsilon$}		(E1);
	\draw[->] (A0) to node[midway,above]		{\small $1-\beta$}		(B0);
	\draw[->] (A0) to node[near start,below]	{\small $\beta$}		(Be);
	\draw[->] (A1) to node[near start,above]	{\small $\beta$}		(Be);
	\draw[->] (A1) to node[midway,below]		{\small $1-\beta$}		(B1);
\end{tikzpicture}
\label{fig:bec_bsc_channel_b}
}
\caption{System model for the wiretap channel with BEC/BSC sources.}
\label{fig:bec_bsc_channel}
\end{figure}

Consider the communication system depicted in Fig.~\ref{fig:bec_bsc_channel}.
The main channel consists of a noiseless link from Alice to Bob and a binary symmetric channel (BSC) with crossover probability $\zeta\in\left[0,\tfrac12\right]$ from Alice to Eve (see Fig.~\ref{fig:bec_bsc_channel_a}).
Additionally, the three nodes have access to correlated sources; in particular, Alice observes a binary uniformly distributed source, i.e., $A\sim\cB\!\left(\tfrac12\right)$, which is the input of two parallel channels as shown in Fig.~\ref{fig:bec_bsc_channel_b}. Bob observes the output of a binary erasure channel (BEC) with erasure probability $\beta\in[0,1]$, and Eve, a BSC with crossover probability $\epsilon\in\left[0,\tfrac12\right]$.
For simplicity, we assume $\eta=1$ in the sequel.

\vspace{1mm}
\begin{remark}\label{rk:bec_bsc_channel_prop}
The sources $(A,B,E)$ satisfy different properties according to the values of the parameters $(\beta,\epsilon)$~\cite{nair_isit_2009}, specifically:
\begin{itemize}
 \item if $0\leq\beta < 2\epsilon$, $E$ is a \emph{degraded} version of $B$, i.e., $A\mkv B\mkv E$;
 \item if $2\epsilon\leq\beta < 4\epsilon(1-\epsilon)$, $B$ is \emph{less noisy} than $E$, i.e., $B\lessnoisy{A}E$; and,
 \item if $4\epsilon(1-\epsilon)\leq\beta < h_2(\epsilon)$, $B$ is \emph{more capable} than $E$.
\end{itemize}
\end{remark}

\subsection{Performance of the Coding Scheme}

The following proposition provides a simple expression of the inner bound from Theorem~\ref{th:digital-key}. The expression is obtained by simplifying the maximization process of the input distribution, and thus it might not be optimal. However, this suffices to show the higher rates achieved by this scheme as we see later.
\vspace*{1mm}
\begin{proposition}
\label{prop:bec_bsc_sep}
The tuple $(\eta=1,R_k)\in\cR_\text{in}$ if there exist $u,v,q\in\big[0,\tfrac12\big]$ such that:
\begin{subequations}\label{eq:bec_bsc_inner_sep}
\begin{align}
&R_k \leq (1-\beta)\big[h_2(v* u) -h_2(v)\big] -h_2(v* u*\epsilon) \nonumber\\
 &\qquad +h_2(v*\epsilon)  +h_2(\zeta) +h_2(q) -h_2(\zeta* q)\,, \\
 &\textnormal{subject to }\ \beta\big[1 -h_2(v* u)\big] \leq 1 -h_2(q)\,. \label{eq:bec_bsc_inner_sep_b}
\end{align}
\end{subequations}
\end{proposition}
\begin{IEEEproof}
The bound~\eqref{eq:bec_bsc_inner_sep} is directly calculated from~\eqref{eq:th1-main_result_rate} and~\eqref{eq:th1-main_result1} with the following choice of input random variables: $T=X$, $Q=X\oplus Q'$, $V=A\oplus V'$, and $U=V\oplus U'$. Here, $X\sim\cB\!\left(\tfrac12\right)$, $Q'\sim\cB(q)$, $V'\sim\cB(v)$, and $U'\sim\cB(u)$, and each random variable is independent of each other and $(A,B,E)$.
The condition~\eqref{eq:th1-main_result2} in the inner bound becomes redundant with the mentioned choice of input distribution.
\end{IEEEproof}
\vspace*{1mm}

As previously mentioned, we provide next the inner bound presented in~\cite[Thm.~4]{khisti_secret-key_2012}\footnote{%
Theorem 4 from~\cite{khisti_secret-key_2012} is actually a capacity result assuming that $A\mkv B\mkv E$ and $X\mkv Y\mkv Z$. In our present example, only the second Markov chain holds independently of the value of the parameters $\beta$ and $\epsilon$, but this does not invalidate the use of the inner bound.%
} as a means of comparison. This inner bound is similar to Theorem~\ref{th:digital-key} but with only one layer of description for the source $A$; thus, its achievable region is denoted $\cR_\text{in-1L}$.

\vspace*{1mm}
\begin{proposition}[{\hspace{1sp}\cite[Thm. 4]{khisti_secret-key_2012}}]
\label{prop:bec_bsc_sep_1l}
The tuple $(\eta=1,R_k)\in\cR_\text{in-1L}$ if and only if
\begin{equation}
 R_k \leq \big[ h_2(\epsilon) -\beta \big]^+ +h_2(\zeta)\,. \label{eq:bec_bsc_inner_sep_1l}
\end{equation}
\end{proposition}
\begin{IEEEproof}
See Appendix~\ref{sec:Proof-BEC_BSC_1L}.
\end{IEEEproof}

\vspace*{1mm}
\begin{remark}
Proposition~\ref{prop:bec_bsc_sep_1l} is a special case of Proposition~\ref{prop:bec_bsc_sep} with $u=q=\tfrac12$, and $v=0$ or $v=\tfrac12$.
As mentioned in Remark~\ref{rk:inner_sep_prev_results}, the inner bound~\cite[Thm.~4]{khisti_secret-key_2012} is a special case of our Theorem~\ref{th:digital-key} with $U=\emptyset$ (thus $u=\tfrac12$).
Moreover, given that in this model the Markov chain $X\mkv Y\mkv Z$ holds, the channel codebook of Proposition~\ref{prop:bec_bsc_sep_1l} has only one layer (thus $q=\tfrac12$).
On the other hand, there are two layers of description in Proposition~\ref{prop:bec_bsc_sep}, and whenever $U\neq\emptyset$ (i.e., $u<\tfrac12$), we have that $Q\neq\emptyset$ (i.e., $q<\tfrac12$).
This relationship is determined by~\eqref{eq:bec_bsc_inner_sep_b}.
\end{remark}
\vspace*{1mm}

\begin{figure}
\centering
\pgfplotsset{every axis/.append style={font=\footnotesize}}
\begin{tikzpicture}
\begin{axis}[
  width=.6\columnwidth,
  scale only axis,
%
  xmin=0,xmax=1,
  ymin=0,ymax=.4,
  grid=both, 
  xlabel=$\beta$,
  ylabel={$R_k$ [bits]},
  ylabel near ticks,
  legend cell align=left,
  legend style={legend pos=north east, font=\scriptsize},
  reverse legend,
  legend entries = { Proposition~\ref{prop:bec_bsc_sep_1l},
		     Proposition~\ref{prop:bec_bsc_sep}
		    },
]
\addplot +[mark=none, black, dashed] coordinates {(2*0.05,0) (2*0.05,1)}; 
\node at (axis cs:0.05,0.05) {\scriptsize A};
\draw[<-,thick] (axis cs:0,0.05) -- (axis cs:0.03,0.05);
\draw[->,thick] (axis cs:0.07,0.05) -- (axis cs:0.1,0.05);
\addplot +[mark=none, black, dashed] coordinates {(4*0.05*(1-0.05),0) (4*0.05*(1-0.05),1)}; 
\node at (axis cs:0.145,0.05) {\scriptsize B};
\draw[<-,thick] (axis cs:0.1,0.05) -- (axis cs:0.125,0.05);
\draw[->,thick] (axis cs:0.165,0.05) -- (axis cs:0.19,0.05);
\addplot +[mark=none, black, dashed] coordinates {(0.2864,0) (0.2864,1)}; 
\node at (axis cs:0.2382,0.05) {\scriptsize C};
\draw[<-,thick] (axis cs:0.19,0.05) -- (axis cs:0.22,0.05);
\draw[->,thick] (axis cs:0.2564,0.05) -- (axis cs:0.2864,0.05);
\addplot +[mark=none, black, dashed] coordinates {(0,.08079) (1,.08079)};
%
\addplot +[mark options= { scale=1 }, mark repeat=200, very thick]						
	table[x=beta,y=rate_1l,col sep=comma] {img/data_bec-bsc.csv};
\addplot +[mark options= { scale=.8, solid }, mark repeat=200, mark phase = 100, very thick]
	table[x=beta,y=rate_2l,col sep=comma] {img/data_bec-bsc.csv};
\end{axis}
\end{tikzpicture}
\caption{Achievable secret key rates for the wiretap channel with BEC/BSC sources, with $\zeta=0.01$ and $\epsilon=0.05$. %
In region A, $A\mkv B\mkv E$, while in region B, $B\lessnoisy{A}E$, and finally in region C, $B$ is more capable than $E$. %
The horizontal dotted line corresponds to the secrecy capacity of the main channel, i.e., $h_2(\zeta)$. %
\label{fig:bec_bsc_channel_rate}}
\end{figure}
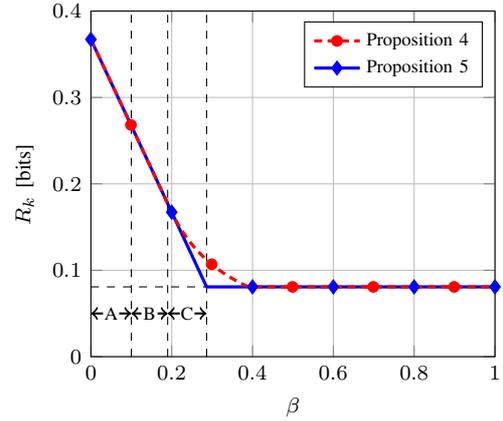

We perform numerical optimization of the bound~\eqref{eq:bec_bsc_inner_sep} for different values of $\beta$ while fixing $\zeta=0.01$ and $\epsilon=0.05$; the results are shown in Fig.~\ref{fig:bec_bsc_channel_rate} along with the bound~\eqref{eq:bec_bsc_inner_sep_1l}.
We see in the figure the advantage of having two layers of description for the source $A$. Our proposed scheme, Proposition~\ref{prop:bec_bsc_sep}, attains higher secret key rates than the scheme with only one layer of description (Proposition~\ref{prop:bec_bsc_sep_1l}) for intermediate values of $\beta$. It is in this regime, when the source $B$ is no longer \emph{less noisy} than $E$, that two layers of description are needed.

\section{Proof of Theorem~\ref{th:digital-key}}
\label{sec:proof-digital}

We begin by presenting a high-level description of the coding strategy before properly developing the proof.
In this scheme, the secret key is learned by extracting common bits from the correlated sources and from exchanging other bits through the noisy channel.
In particular, Alice compresses the source observation $A^n$ using a two-layer source codebook (determined by $U^n$ and $V^n$).
Alice then transmits the corresponding bin indices $r_1$ and $(r_2,r_p)$ to Bob with the aid of a code for the wiretap channel.
Using his side-information $B^n$, Bob recovers the codewords $U^n$ and $V^n$ and he further obtains the bin indices $(r_2,k_1)$, where $k_1$ is independent of $r_p$ provided that the conditions of Lemma~\ref{lem:lem_fano_eve} (Section~\ref{ssec:key_leakage}) are met.
The key is finally generated using bits from the indices $k_1$ and $k_2$, where the latter was sent over the noisy channel along with $r_1$, $r_2$, and $r_p$.
We provide a detailed proof in the following.

\subsection{Codebook Generation}
\label{ssec:key_codebook}

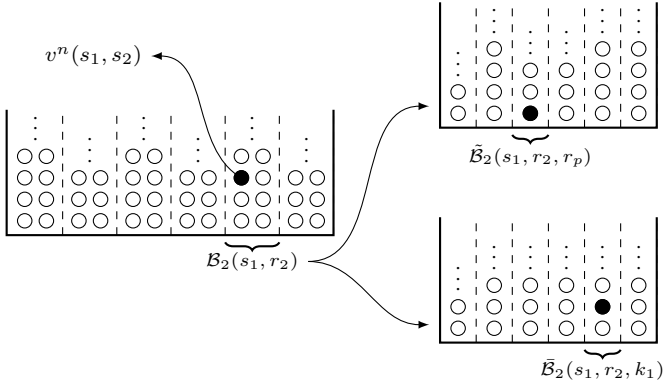
\begin{figure}[t!]
\centering
\begin{tikzpicture}[font=\footnotesize, scale=.95] 
\foreach \i in {0,...,5} { %
  \pgfmathsetmacro{\numcods}{floor(rand+3)}
  \foreach \j in {0,...,\numcods} { %
    \draw (.25+\i*0.75,1.2+\j*0.3) circle (0.1);
    \draw (.55+\i*0.75,1.2+\j*0.3) circle (0.1);
  }
  \draw (.4+\i*0.75,1.2+\numcods*0.3+0.5) node {$\vdots$};
  \draw[dashed] (.775+\i*0.75,1) -- (.775+\i*0.75,2.75);
}
\draw[thick] (0,2.75) -- (0,1) -- (4.525,1) -- (4.525,2.75);
\draw [thick,decorate,decoration={brace,amplitude=3pt}] (3.775,.9) -- (3.025,.9) node (B) [midway,below] {\scriptsize $\cB_2(s_1,r_2)$};
\draw[fill] (.25+4*0.75,1.2+2*0.3) circle (0.1);
%
\foreach \i in {0,...,5} { %
  \pgfmathsetmacro{\numcods}{floor(rand+2.75)}
  \foreach \j in {0,...,\numcods} { %
    \draw (6.25+\i*0.5,2.7+\j*0.3) circle (0.1);
  }
  \draw (6.25+\i*0.5,2.7+\numcods*0.3+0.5) node {$\vdots$};
  \draw[dashed] (6.5+\i*0.5,2.5) -- (6.5+\i*0.5,4.25);
}
\draw[thick] (6,4.25) -- (6,2.5) -- (9,2.5) -- (9,4.25);
\draw [thick,decorate,decoration={brace,amplitude=3pt}] (7.5,2.4) -- (7,2.4) node [midway,below] {\scriptsize $\tilde{\cB}_2(s_1,r_2,r_p)$};
\draw[fill] (6.25+2*0.5,2.7+0*0.3) circle (0.1);
%
\foreach \i in {0,...,5} { %
  \pgfmathsetmacro{\numcods}{floor(rand+2.75)}
  \foreach \j in {0,...,\numcods} { %
    \draw (6.25+\i*0.5,-.3+\j*0.3) circle (0.1);
  }
  \draw (6.25+\i*0.5,-.3+\numcods*0.3+0.5) node {$\vdots$};
  \draw[dashed] (6.5+\i*0.5,-.5) -- (6.5+\i*0.5,1.25);
}
\draw[thick] (6,1.25) -- (6,-.5) -- (9,-.5) -- (9,1.25);
\draw [thick,decorate,decoration={brace,amplitude=3pt}] (8.5,-.6) -- (8,-.6) node [midway,below] {\scriptsize $\bar{\cB}_2(s_1,r_2,k_1)$};
\draw[fill] (6.25+4*0.5,-.3+1*0.3) circle (0.1);
%
\draw [-latex] (B) to [out=0,in=180] (5.85,2.8);
\draw [-latex] (B) to [out=0,in=180] (5.85,-.25);
%
\draw [-latex] (.25+4*0.75,1.2+2*0.3) to [out=150,in=0] (2,3.5) node[left] {$v^n(s_1,s_2)$};
\end{tikzpicture}
\caption{Multiple binning stages of the codewords $v^n(s_1,s_2)$, where each circle represents a codeword. 
The fact that a codeword belongs to different sub-bins $\tilde{\cB}_2$ and $\bar{\cB}_2$ is shown through a black circle, which depicts the same codeword.
\label{fig:binning}}
\end{figure}

Let us define the quantity
\begin{equation}
 R_f < (\eta+\epsilon) \IC{T}{Z}{Q} -\epsilon_f\,,
 \label{eq:th1-cond_rf}
\end{equation}
and fix the following joint probability distribution:
\begin{multline}
 p(qtxyzuvabe) = \\ p(q\vert t)p(tx)p(yz|x) p(u\vert v)p(v\vert a)p(abe)\,.
 \label{eq:th1-pmf}
\end{multline}
Then, proceed as follows:
\begin{enumerate}
 \item Randomly pick $2^{nS_1}$ sequences $u^n(s_1)$ from $\typ{n}{U}$ and divide them into $2^{nR_1}$ equal-size bins $\cB_1(r_1)$, $r_1\in[1:2^{nR_1}]$.
 \item For each codeword $u^n(s_1)$, randomly pick $2^{nS_2}$ sequences $v^n(s_1,s_2)$ from $\typ{n}{V|u^n(s_1)}$\footnote{%
 As a matter of fact, the sequences $v^n(s_1,s_2)$ should be chosen from $\typc{n}{V|u^n(s_1)}$, $\delta<\delta'$, in order to assure that $(u^n(s_1), v^n(s_1,s_2))\in \typc{n}{UV}$ (see e.g. Conditional Typicality Lemma~\cite{gamal_network_2011}). This remark also applies to the generation of the channel codewords $q^m(\cdot)$ and $t^m(\cdot)$ in this part of the proof. However, we omit this detail throughout the proof to simplify the notation and ease the reading.
 } and divide them into $2^{nR_2}$ equal-size bins $\cB_2(s_1,r_2)$, $r_2\in[1:2^{nR_2}]$. Furthermore, distribute the sequences inside each bin in two different types of sub-bin:
 \begin{itemize}
  \item $2^{nR_p}$ equal-size sub-bins $\tilde{\cB}_2(s_1,r_2,r_p)$, $r_p\in[1:2^{nR_p}]$; and,
  \item $2^{nR_{k_1}}$ equal-size sub-bins $\bar{\cB}_2(s_1,r_2,k_1)$, $k_1\in[1:2^{nR_{k_1}}]$.
 \end{itemize}
 Note that a sequence $v^n(s_1,s_2)$ belongs to sub-bins $\tilde{\cB}_2(s_1,r_2,r_p)$ and $\bar{\cB}_2(s_1,r_2,k_1)$ where $r_p$ and $k_1$ are independent. See Fig.~\ref{fig:binning} for a schematic representation.
 \item Randomly pick $2^{n(R_1+R_2)}$ sequences $q^m(r_1,r_2)$ from $\typ{m}{Q}$.
 \item For each $q^m(r_1,r_2)$, randomly pick $2^{n(R_p+R_{k_2}+R_f)}$ sequences $t^m(r_1,r_2,r_p,k_2,r_f)$ from $\typ{m}{T|q^m(r_1,r_2)}$.
 \item Randomly distribute the set of $2^{n(R_{k_1}+R_{k_2})}$ indices $(k_1,k_2)$ into $2^{nR_k}$ equal-size bins $\cB_k(k)$, $k\in[1:2^{nR_k}]$.
\end{enumerate}

\subsection{Encoding}
\label{ssec:key_encoding}

Given a sequence $a^n$, and the indices $k_2$ and $r_f$ chosen uniformly at random in $[1:2^{nR_{k_2}}]$ and $[1:2^{nR_f}]$, the encoder proceeds as follows:
\begin{enumerate}
 \item It looks for an index $s_1 \equiv\hat{s}_1$ such that $(u^n(\hat{s}_1),a^n)\in\typc{n}{UA}$.
 If more than one index is found, choose one uniformly at random among them, whereas if there is no such index, choose one uniformly at random in $[1:2^{nS_1}]$.
 The probability of not finding such an index is arbitrarily small as $n\rightarrow\infty$ if $\delta^\prime<\epsilon_1$ and
  \begin{equation}
  S_1 > \I{U}{A} +\epsilon_1\,.
  \end{equation}
 \item Then, it looks for an index $s_2 \equiv\hat{s}_2$ such that $(v^n(s_1,\hat{s}_2),a^n)\in\typc{n}{VA|u^n(s_1)}$.
 If more than one index is found, choose one uniformly at random among them, whereas if there is no such index, choose one uniformly at random in $[1:2^{nS_2}]$.
 The probability of not finding such an index is arbitrarily small as $n\rightarrow\infty$ if $\delta^\prime<\epsilon_2$ and
  \begin{equation}
  S_2 > \IC{V}{A}{U} +\epsilon_2\,. \label{eq:th1-enc2}
  \end{equation}
 \item Let $\cB_1(r_1)$ and $\tilde{\cB}_2(s_1,r_2,r_p)$ be the bins of $u^n(s_1)$ and $v^n(s_1,s_2)$, respectively.
 \item The encoder selects the codeword $t^m(r_1,r_2,r_p,k_2,r_f)$. It then transmits the associated jointly typical sequence $x^m \sim \prod_{i=1}^m p(x_i\vert t_i(r_1,r_2,r_p,k_2,r_f))$, generated on the fly.
\end{enumerate}

\subsection{Decoding}

Given a sequence $b^n$ and the channel output $y^m$, the decoder proceeds as follows:
\begin{enumerate}
 \item It starts by looking for the unique set of indices $(r_1,r_2,r_p,k_2,r_f) \equiv (\hat{r}_1,\hat{r}_2, \hat{r}_p,\hat{k}_2, \hat{r}_f)$ such that
  \begin{equation*}
  \big(q^m(\hat{r}_1,\hat{r}_2), t^m(\hat{r}_1,\hat{r}_2, \hat{r}_p,\hat{k}_2, \hat{r}_f), y^m\big)\in\typ{m}{QTY}\,.
  \end{equation*}
 The probability of error in decoding can be made arbitrarily small as $(n,m)\rightarrow\infty$ provided that
  \begin{align*}
  R_1 +R_2 +R_p +R_{k_2} +R_f &< (\eta+\epsilon)\I{T}{Y} -\delta\,,\\
            R_p +R_{k_2} +R_f &< (\eta+\epsilon)\IC{T}{Y}{Q} -\delta\,.
  \end{align*}
 \item The decoder looks for the unique index $s_1 \equiv \hat{s}_1$ such that $u^n(\hat{s}_1)\in \cB_1(r_1)$ and $(u^n(\hat{s}_1),b^n) \in\typ{n}{UB}$. The probability of error in decoding can be made arbitrarily small as $n\rightarrow\infty$ provided that
  \begin{equation}
  S_1 -R_1 < \I{U}{B} -\delta\,.
  \end{equation}
 \item Then, it looks for the unique index $s_2 \equiv \hat{s}_2$ such that $v^n(s_1,\hat{s}_2)\in \tilde{\cB}_2(s_1,r_2,r_p)$ and $(v^n(s_1,\hat{s}_2), b^n)\in\typ{n}{VB|u^n(s_1)}$. The probability of error in decoding can be made arbitrarily small as $n\rightarrow\infty$ provided that
  \begin{equation}
  S_2 -R_2 -R_p < \IC{V}{B}{U} -\delta\,.
  \end{equation}
\end{enumerate}

\subsection{Key Generation}

According to the preceding steps and with increasing high probability as $(n,m)\to\infty$, Bob correctly decodes the index $k_2$ and both Alice and Bob possess the same sequence $v^n(s_1,s_2)\in\bar{\cB}_2(s_1,r_2,k_1)$. Therefore, they both agree on the same secret key $k$, which is the bin where the pair $(k_1,k_2)$ belongs, i.e., $(k_1,k_2)\in\cB_k(k)$.

\subsection{Key Uniformity}

Consider the following chain of inequalities:
\begin{subequations} \label{eq:th1-keyunif_1}
\begin{align}
 \HC{K}{\cC} &= \HC{K_1 K_2}{\cC} -\HC{K_1 K_2}{\cC K} \\
 &\geq \HC{K_1}{\cC} +nR_{k_2} -n(R_{k_1} +R_{k_2} -R_{k}) \label{eq:th1-keyunif_1b} \\
 &\geq \HC{K_1}{\cC U^n} -n(R_{k_1} -R_{k}) \\
 &= \HC{V^n}{\cC U^n} -\HC{V^n}{\cC U^n K_1} -n(R_{k_1} -R_{k}) \label{eq:th1-keyunif_1d} \\
 &\geq \HC{V^n}{\cC U^n} -n(S_2-R_{k_1}) -n(R_{k_1} -R_{k})\,, \label{eq:th1-keyunif_1e}
\end{align}
\end{subequations}
where
\begin{itemize}
 \item \eqref{eq:th1-keyunif_1b} follows from $K_2$ being chosen uniformly in $[1:2^{nR_{k_2}}]$ and independently of $K_1$, and that there are $2^{n(R_{k_1} +R_{k_2} -R_{k})}$ pairs $(K_1,K_2)$ in each bin $K$;
 \item \eqref{eq:th1-keyunif_1d} is due to $K_1$ being a function of $(V^n,\cC)$; and,
 \item \eqref{eq:th1-keyunif_1e} is due to the number of sequences $V^n$ associated with sub-bin index $K_1$ being $2^{n(S_2-R_{k_1})}$, i.e., $\log \sum_{r_2} |\bar{\cB}_2(s_1,r_2,k_1)| = n(S_2-R_{k_1})$.
\end{itemize}

Before continuing the analysis, we introduce the random variable $\Upsilon$, such that
\begin{equation}
\Upsilon \triangleq \ind{(U^n,A^n) \in \typ{n}{UA}}\,.
\end{equation}
Moreover, in order to improve readability, we drop the index from the codeword $U^n$, and thus the codebook $\cC$ is composed of: $U^n\in\typ{n}{U}$ and $V^n(s)\in\typ{n}{V|U^n}$, where $s\in\cS\triangleq[1:2^{nS_2}]$.
Finally, we note that, conditioned on the codebook $\cC$, the entropy of $V^n$ is given by the entropy of its index $S$.
Therefore,
\begin{subequations} \label{eq:th1-keyunif_2}
\begin{align}
 \HC{V^n}{\cC U^n} &= \HC{S}{\cC U^n} \\
 &\geq \HC{S}{\cC U^n \Upsilon} \\
 &\geq \HC{S}{\cC U^n, \Upsilon=1} (1-\epsilon)\,,
\end{align}
\end{subequations}
where the last step is due to $\PR{\Upsilon=1}\geq 1-\epsilon$.

Now, for a specific codebook $\cC=\mathsf{c}_n$ (which determines the codeword $U^n=u^n$), let us define the random variable $S_c$ with distribution
\begin{equation}
 p_{S_c} \triangleq p_{S\vert \cC=\mathsf{c}_n, U^n=u^n, \Upsilon=1}\,.
\end{equation}
Therefore,
\begin{equation}
\H{S_c} = \HC{S}{\cC=\mathsf{c}_n, U^n=u^n, \Upsilon=1}\,, 
\end{equation}
and
\begin{subequations} \label{eq:th1-keyunif_3}
\begin{align}
 \HC{S}{\cC U^n, \Upsilon=1} &= \bE_{\cC} \big[ \H{S_c} \big] \\
 &= \sum_{s\in\cS} \bE_{\cC} \big[ -p_{S_c}(s) \log p_{S_c}(s) \big] \\
 &= |\cS|  \, \bE_{\cC}\big[ -p_{S_c}(1) \log p_{S_c}(1) \big]\,,
\end{align}
\end{subequations}
where the last step is due to the symmetry of the random codebook generation and encoding procedure, i.e., the probability $p_{S_c}$ is independent of the specific value of the index.
This is addressed in the following lemma.

\vspace{1mm}
\begin{lemma}\label{lem:lem_bnd_ps}
Let $\varepsilon_1, \varepsilon_2, \xi>0$ and let $\chi$ be a function of the codebook $\mathsf{c}_n$ defined as
\begin{equation}
 \chi(\mathsf{c}_n) \triangleq \ind{ \big| p_{S_c}(1) -|\cS|^{-1} \big| \geq \varepsilon_1\, |\cS|^{-1} }\,. 
\end{equation}
Then, $\PR{ \chi(\cC)=1 }\leq \varepsilon_2$ for large $n$ if $S_2<\H{A}-\xi$.
\end{lemma}
\begin{IEEEproof}
This lemma is similar to the one introduced in~\cite[Lemma~5]{bassi_wiretap_2015} and its proof is reproduced in Appendix~\ref{sec:Proof-Lemma-Bound_ps} for completeness.
\end{IEEEproof}
\vspace{1mm}

Using the previous lemma we may continue~\eqref{eq:th1-keyunif_3},
\begin{subequations} \label{eq:th1-keyunif_4}
\begin{align}
\MoveEqLeft[1]
 \HC{S}{\cC U^n, \Upsilon=1} \nonumber\\
 &\geq |\cS|  \, \bE_{\cC}\big[ -p_{S_c}(1) \log p_{S_c}(1) \mid \chi(\cC)=0 \big] (1-\varepsilon_2) \\
 &\geq (1-\varepsilon_1) \big[ \log |\cS| -\log(1+\varepsilon_1) \big] (1-\varepsilon_2) \\
 &\geq n(S_2 -\varepsilon')\,,
\end{align}
\end{subequations}
for some $\varepsilon'>0$. Putting together~\eqref{eq:th1-keyunif_1}, \eqref{eq:th1-keyunif_2}, and~\eqref{eq:th1-keyunif_4}, we obtain
\begin{align}
  \HC{K}{\cC} \geq n(S_2 -\varepsilon') (1-\epsilon) -n(S_2 -R_k) \geq n(R_k -\epsilon')\,,
\end{align}
for some $\epsilon'>0$.
Finally, the uniformity of the keys, as defined in~\eqref{eq:def_sk_unif}, averaged over all codebooks is
\begin{align}
 \bE[\mathsf{U}_k(\cC)] = nR_k -\HC{K}{\cC} \leq n\epsilon'\,,
\end{align}
and thus the key is asymptotically uniform.

\vspace{1mm}
\begin{remark}\label{rk:uniform_index}
It is worth noting that the preceding steps show that the probability of $V^n$ is almost uniformly distributed on the codebook,
\begin{equation}
 \HC{V^n}{\cC U^n} \geq n(S_2 -\varepsilon') (1-\epsilon)\,,
\end{equation}
which follows from~\eqref{eq:th1-keyunif_2} and~\eqref{eq:th1-keyunif_4}. A lower bound on $\HC{U^n}{\cC}$ may be obtained using a similar analysis.
Given that the sequences $U^n$ and $V^n$ are divided randomly and independently on equal-size bins and sub-bins, the bin and sub-bin indices (e.g. $r_p$) are also distributed almost uniformly on their respective sets.
\end{remark}

\subsection{Key Leakage}
\label{ssec:key_leakage}

We may first relate the entropy of $K$ to that of $(K_1,K_2)$ as in~\eqref{eq:th1-keyunif_1},
\begin{subequations}\label{eq:th1-keyleak_0}
\begin{align}
\MoveEqLeft[1]
 \HC{K}{\cC E^n Z^m} \nonumber\\
 &= \HC{K_1 K_2}{\cC E^n Z^m} -\HC{K_1 K_2}{\cC E^n Z^m K} \\
 &\geq \HC{K_1 K_2}{\cC E^n Z^m} -n(R_{k_1} +R_{k_2} -R_{k})\,.
\end{align}
\end{subequations}
Then, consider the following chain of inequalities:
\begin{subequations}\label{eq:th1-keyleak_1}
\begin{align}
\MoveEqLeft[1]
 \HC{K_1 K_2}{\cC E^n Z^m} \nonumber\\
 &\geq \HC{K_1 K_2}{\cC E^n Z^m r_1 r_2} \nonumber\\
 &= \HC{K_2 U^n V^n}{\cC E^n Z^m r_1 r_2} \nonumber\\
 &\quad -\HC{U^n V^n}{\cC E^n Z^m r_1 r_2 K_1 K_2} \label{eq:th1-keyleak_1b}\\
 &\geq \HC{K_2 U^n V^n}{\cC E^n Z^m r_1 r_2} -\HC{U^n}{\cC E^n r_1} \nonumber\\
 &\quad -\HC{V^n}{\cC E^n Z^m U^n r_2 K_1 K_2} \nonumber\\
 &\geq \HC{K_2 U^n V^n}{\cC E^n Z^m r_1 r_2} -2n\epsilon_n \label{eq:th1-keyleak_1d}\\
 &= \HC{K_2 U^n V^n A^n}{\cC E^n Z^m r_1 r_2} \nonumber\\
 &\quad -\HC{A^n}{\cC E^n Z^m U^n V^n K_2} -2n\epsilon_n \label{eq:th1-keyleak_1e} \displaybreak[2]\\
 &\geq \HC{K_2 A^n}{\cC E^n Z^m r_1 r_2} -\HC{A^n}{U^n V^n E^n} -2n\epsilon_n \nonumber\displaybreak[2]\\
 &\geq \HC{A^n}{\cC E^n Z^m r_1 r_2 K_2} +\HC{K_2}{\cC E^n Z^m r_1 r_2} \nonumber\\
 &\quad -n[\HC{A}{U V E} +2\epsilon_n] \nonumber\displaybreak[2]\\
 &= {\underbrace{ \HC{A^n}{\cC E^n Z^m r_1 r_2 r_p K_2} }_{\triangleq E_s}} \nonumber\\
 &\quad + {\underbrace{ \IC{A^n}{r_p}{\cC E^n Z^m r_1 r_2 K_2} +\HC{K_2}{\cC E^n Z^m r_1 r_2} }_{\triangleq E_c}} \nonumber\\
 &\quad -n[\HC{A}{U V E} +2\epsilon_n]\,,
\end{align}
\end{subequations}
where
\begin{itemize}
 \item \eqref{eq:th1-keyleak_1b} is due to $K_1$ being a function of $(V^n,\cC)$; 
 \item \eqref{eq:th1-keyleak_1d} follows from Lemma~\ref{lem:lem_fano_eve} below; and,
 \item \eqref{eq:th1-keyleak_1e} is due to $(r_1,r_2)$ being functions of $(U^n,V^n,\cC)$.
\end{itemize}

\vspace{1mm}
\begin{lemma}\label{lem:lem_fano_eve}
Let $\epsilon_n, \delta, \delta' ,\varepsilon_1>0$, then, given the codebook generation and encoding procedure of the scheme,
\begin{subequations}\label{eq:lem_fano_eve}
\begin{equation}
 \HC{U^n}{\cC E^n r_1} \leq n\epsilon_n
\end{equation}
if $S_1 -R_1 < \I{U}{E} -\delta$, and
\begin{equation}
 \HC{V^n}{\cC E^n Z^m U^n r_2 K_1 K_2} \leq n\epsilon_n \label{eq:lem_fano_eve2}
\end{equation}
\end{subequations}
if $S_2 -R_2 -R_{k_1} +R_f < \IC{V}{E}{U} + (\eta+\epsilon) \IC{T}{Z}{Q} -\delta'$ and $R_p +R_f >(\eta+\epsilon)\IC{T}{Z}{Q} +\varepsilon_1$.
\end{lemma}
\begin{IEEEproof}
See Appendix~\ref{sec:Proof-Lemma-Fano_Eve}.
\end{IEEEproof}
\vspace{1mm}

In the last step of~\eqref{eq:th1-keyleak_1}, we split up the equivocation into two parts as in~\cite{villard_secure_2014}.
The ``source'' term $E_s$ writes:
\begin{subequations}\label{eq:th1-keyleak_2}
\begin{align}
E_s &= \HC{A^n}{\cC E^n r_1 r_2 r_p} \label{eq:th1-keyleak_2a} \\
    &= \HC{A^n r_2 r_p}{\cC E^n r_1} -\HC{r_2 r_p}{\cC E^n r_1} \\
    &= \HC{A^n}{\cC E^n r_1} +\HC{r_2 r_p}{\cC A^n E^n r_1} -\HC{r_2 r_p}{\cC} \nonumber\\
    &\quad +\IC{r_2 r_p}{E^n r_1}{\cC} \\
    &\geq \HC{A^n}{U^n E^n} +\HC{r_p}{\cC A^n E^n r_1 r_2} -n(R_2 +R_p) \nonumber\\
    &\quad +\IC{r_2 r_p}{E^n}{\cC r_1} \label{eq:th1-keyleak_2d}\\
    &\geq n[\HC{A}{U E} -\varepsilon] -n(R_2 +R_p)  +\HC{r_p}{\cC A^n E^n r_1 r_2} \nonumber\\
    &\quad +\IC{r_2 r_p}{E^n}{\cC r_1}\,, \label{eq:th1-keyleak_2e}
\end{align}
\end{subequations}
where
\begin{itemize}
 \item \eqref{eq:th1-keyleak_2a} follows from the Markov chain $(A^n E^n)\mkv (\cC r_1 r_2 r_p)\mkv(K_2 Z^m)$;
 \item \eqref{eq:th1-keyleak_2d} is due to the Markov chain $(A^n E^n) \mkv U^n \mkv (r_1 \cC)$, the fact that the indices $r_2$ and $r_p$ belong to sets of cardinality $2^{nR_2}$ and $2^{nR_p}$, and the non-negativity of mutual information; and,
 \item \eqref{eq:th1-keyleak_2e} stems from the lower bound found on Lemma~\ref{lem:lem_lower_bnd_typ} below.
\end{itemize}

\vspace{1mm}
\begin{lemma}\label{lem:lem_lower_bnd_typ}
Given the codebook generation and encoding procedure of the scheme,
\begin{equation}
 \HC{A^n}{U^n E^n} \geq n [\HC{A}{U E} -\varepsilon]\,.
\end{equation}
\end{lemma}
\begin{IEEEproof}
Using well-known properties of typical sets, we have
\begin{align*}
\HC{A^n}{U^n E^n} &= -\quad \sum_{\mathclap{\forall (u^n a^n e^n)}} \quad\ p(u^n a^n e^n) \log p(a^n\vert u^n e^n) \\
 &\geq -\quad \sum_{\mathclap{(u^n a^n e^n) \in\typ{n}{UAE}}} \quad\ p(u^n a^n e^n) \log p(a^n\vert u^n e^n) \\
 &\geq \qquad \sum_{\mathclap{(u^n a^n e^n) \in\typ{n}{UAE}}} \quad\ p(u^n a^n e^n)\, n[\HC{A}{U E} -\varepsilon^{(1)}] \displaybreak[2]\\
 &\geq (1-\varepsilon^{(2)})\, n[\HC{A}{U E} -\varepsilon^{(1)}] \\
 &\geq n [\HC{A}{U E} -\varepsilon^{(3)}]\,,
\end{align*}
where in the last step we choose $\varepsilon^{(3)}$ large enough to have a lower bound.
\end{IEEEproof}
\vspace{1mm}

On the other hand, the ``channel'' term $E_c$ writes:
\begin{align}
E_c &= \HC{r_p K_2}{\cC E^n Z^m r_1 r_2} -\HC{r_p}{\cC A^n E^n Z^m r_1 r_2 K_2} \nonumber\\
    &= \HC{r_p K_2}{\cC Z^m r_1 r_2} -\IC{r_p K_2}{E^n}{\cC Z^m r_1 r_2} \nonumber\\
    &\quad -\HC{r_p}{\cC A^n E^n Z^m r_1 r_2 K_2}\,. \label{eq:th1-keyleak_3b}
\end{align}
The first term on the r.h.s.~of~\eqref{eq:th1-keyleak_3b} corresponds to the equivocation (of the \emph{private} message, given the \emph{common} message and the output of the channel) in the wiretap channel setting. Following the arguments of~\cite[Sec.~IV]{csiszar_broadcast_1978} and~\cite[Sec.~2.3]{liang_its_2008}, together with constraint~\eqref{eq:th1-cond_rf} and Remark~\ref{rk:uniform_index}, we can easily prove the following lower bound\footnote{%
Remark~\ref{rk:uniform_index} assures that the indices $r_1$, $r_2$, and $r_p$ are distributed almost uniformly, a condition that is necessary to invoke the result from the wiretap channel setting.}:
\begin{multline}
\HC{r_p K_2}{\cC Z^m r_1 r_2} \\ \geq n \big[R_p +R_{k_2} +R_f -(\eta+\epsilon)\IC{T}{Z}{Q} -\varepsilon'\big]\,, \label{eq:th1-keyleak_4}
\end{multline}
for sufficiently large $n$.

Gathering \eqref{eq:th1-keyleak_0}, \eqref{eq:th1-keyleak_1}, \eqref{eq:th1-keyleak_2}, \eqref{eq:th1-keyleak_3b}, and~\eqref{eq:th1-keyleak_4}, we have that
\begin{align}
\MoveEqLeft[1]
 \HC{K}{\cC E^n Z^m} \nonumber\\
 &\geq n \big[\IC{V}{A}{U E} -R_{k_1} +R_{k} -R_2 +R_f \nonumber\\
 &\quad -(\eta+\epsilon)\IC{T}{Z}{Q} -\varepsilon''\big] +\IC{r_p}{Z^m K_2}{\cC A^n E^n r_1 r_2} \nonumber\\
 &\quad  +\IC{r_2 r_p}{E^n}{\cC r_1} -\IC{r_p K_2}{E^n}{\cC Z^m r_1 r_2}\,, \label{eq:th1-keyleak_5}
\end{align}
for some $\varepsilon''>0$.
We now study the last two multi-letter terms on the r.h.s.~of~\eqref{eq:th1-keyleak_5}:
\begin{subequations}\label{eq:th1-keyleak_6}
\begin{align}
\MoveEqLeft[1]
\IC{r_2 r_p}{E^n}{\cC r_1} -\IC{r_p K_2}{E^n}{\cC Z^m r_1 r_2} \nonumber\\
 &= \IC{r_2 r_p}{E^n}{\cC r_1} -\IC{r_2 r_p Z^m K_2}{E^n}{\cC r_1} \nonumber\\
 &\quad +\IC{r_2 Z^m}{E^n}{\cC r_1} \\
 &= -\IC{Z^m K_2}{E^n}{\cC r_1 r_2 r_p} +\IC{r_2 Z^m}{E^n}{\cC r_1} \\
 &= \IC{r_2 Z^m}{E^n}{\cC r_1} \label{eq:th1-keyleak_6c}\\
 &\geq 0, \label{eq:th1-keyleak_6d}
\end{align}
\end{subequations}
where
\begin{itemize}
 \item \eqref{eq:th1-keyleak_6c} stems from the Markov chain $E^n \mkv (\cC r_1 r_2 r_p) \mkv (Z^m K_2)$; and,
 \item \eqref{eq:th1-keyleak_6d} is due to the non-negativity of mutual information.
\end{itemize}

Inequality~\eqref{eq:th1-keyleak_5} may then be lower bounded as
\begin{subequations}
\begin{align}
\MoveEqLeft[1]
\HC{K}{\cC E^n Z^m} \nonumber\\
 &\geq n \big[ \IC{V}{A}{U E} -R_{k_1} +R_{k} -R_2 +R_f \nonumber\\
 &\quad -(\eta+\epsilon)\IC{T}{Z}{Q} -\varepsilon'' \big] \\
 &\geq n \big( R_{k} -\varepsilon'' \big)\,,
\end{align}
\end{subequations}
where the last inequality holds if
\begin{equation}
 R_{k_1} +R_2 -R_f \leq \IC{V}{A}{U E} -(\eta+\epsilon)\IC{T}{Z}{Q}\,. \label{eq:th1-keyleak_7}
\end{equation}
Finally,
\begin{align*}
\bE[\mathsf{L}_k(\cC)] &= \IC{K}{E^n Z^m}{\cC}\\
&= \HC{K}{\cC} -\HC{K}{\cC E^n Z^m} \\
&\leq n\varepsilon''\,,
\end{align*}
and the key is asymptotically secure.

\subsection{Sufficient Conditions}

Putting all pieces together, we have proved that the proposed scheme allows the legitimate users to agree upon a key of rate $R_k$, while keeping it secret from the eavesdropper if
\begin{align*}
R_1       &\leq S_1\,,                                      \\
R_2       &\leq S_2\,,                                      \\
R_p       &\leq S_2 -R_2\,,                                 \\
R_{k_1}   &\leq S_2 -R_2\,,                                 \\
R_k       &\leq R_{k_1} +R_{k_2}\,,         \displaybreak[2]\\
R_f       &<    (\eta+\epsilon) \IC{T}{Z}{Q} -\epsilon_f\,, \displaybreak[2]\\
S_1       &>    \I{U}{A} +\epsilon_1\,,                     \\
S_2       &>    \IC{V}{A}{U} +\epsilon_2\,,                 \displaybreak[2]\\
            R_p +R_{k_2} +R_f &< (\eta+\epsilon) \I{T}{Y} -\delta -R_1 -R_2\,,     \\
          R_p +R_{k_2} +R_f &< (\eta+\epsilon) \IC{T}{Y}{Q} -\delta\,, \displaybreak[2]\\
S_1 -R_1                    &< \I{U}{B} -\delta\,,                     \\
S_2 -R_2 -R_p               &< \IC{V}{B}{U} -\delta\,,                 \displaybreak[2]\\
S_2                    &<    \H{A} -\xi\,,                                          \\
S_1 -R_1               &<    \I{U}{E} -\delta\,,                                    \\
S_2 -R_2 -R_{k_1} +R_f &<    \IC{V}{E}{U} +(\eta+\epsilon) \IC{T}{Z}{Q} -\delta'\!, \\
R_p +R_f               &>    (\eta+\epsilon) \IC{T}{Z}{Q} +\varepsilon_1\,,         \\
R_{k_1} +R_2 -R_f      &\leq \IC{V}{A}{UE} -(\eta+\epsilon) \IC{T}{Z}{Q}\,.         
\end{align*}
After applying Fourier-Motzkin elimination to this set of inequalities and taking $(n,m)\rightarrow\infty$, we obtain~\eqref{eq:th1-main_result_rate}
subject to the conditions~\eqref{eq:th1-main_result} and
\begin{subequations}\label{eq:th1-redundant_cond}
\begin{align}
 \IC{T}{Z}{Q}                &\leq \IC{T}{Y}{Q}\,,   \label{eq:th1-redundant_cond0}\\
 \IC{U}{A}{E}                &\leq \eta\,\I{Q}{Y}\,, \label{eq:th1-redundant_cond1}\\
 \IC{U}{A}{E} +\IC{V}{A}{UB} &\leq \eta\,\I{T}{Y}\,. \label{eq:th1-redundant_cond2}
\end{align}
\end{subequations}
The achievable region $\cR_\text{in}$ is the convex hull of the union of this region over all joint probability distributions $p\in\cP$, where the elements of $\cP$ are defined in~\eqref{eq:th1-pmf}.
We show next that the same achievable region is obtained by the convex hull of the union of the region defined by~\eqref{eq:th1-main_result_rate} and~\eqref{eq:th1-main_result} over all $p\in\cP$; therefore we prefer this more compact version.

The conditions~\eqref{eq:th1-redundant_cond1} and~\eqref{eq:th1-redundant_cond2} are redundant whenever $\I{U}{B}\leq\I{U}{E}$, whereas if $(U,V)\sim p(u,v)$ are such that $\I{U}{B}>\I{U}{E}$ while satisfying~\eqref{eq:th1-main_result} and~\eqref{eq:th1-redundant_cond}, we see that
\begin{align*}
\MoveEqLeft[1]
 \I{V}{B} -\I{V}{E} \nonumber\\
 &= \IC{V}{B}{U} -\IC{V}{E}{U} +\I{U}{B} -\I{U}{E} \\
 &> \IC{V}{B}{U} -\IC{V}{E}{U}\,.
\end{align*}
This implies that a larger achievable secret key rate is obtained with $U=\emptyset$ and $V\sim p(v)=\sum_u p(u,v)$, which still satisfies~\eqref{eq:th1-main_result} and~\eqref{eq:th1-redundant_cond}.
Similarly, we see that if $(Q,T)$ are such that $\IC{T}{Z}{Q}> \IC{T}{Y}{Q}$ while satisfying~\eqref{eq:th1-main_result} and~\eqref{eq:th1-redundant_cond},
\begin{multline*}
 \eta\big[ \IC{T}{Y}{Q} -\IC{T}{Z}{Q} \big] +\IC{V}{B}{U} -\IC{V}{E}{U} \\ < \IC{V}{B}{U} -\IC{V}{E}{U}\,.
\end{multline*}
This implies that the achievable secret key rate is increased by choosing $Q=T$, which still satisfies~\eqref{eq:th1-main_result} and~\eqref{eq:th1-redundant_cond}.
%
%
Therefore, the conditions~\eqref{eq:th1-redundant_cond} are redundant after the maximization and may be discarded.

We have shown thus far that, \emph{averaged} over all possible codebooks, the probability of error~\eqref{eq:def_sk_rel}, the key leakage~\eqref{eq:def_sk_leak}, and the uniformity of the keys~\eqref{eq:def_sk_unif} become negligible as $(n,m)\to\infty$ if the conditions~\eqref{eq:th1-main_result_rate} and~\eqref{eq:th1-main_result} hold true. Nonetheless, by applying the selection lemma~\cite[Lemma 2.2]{bloch_physical_2011}, we may conclude that there exists a \emph{specific} sequence of codebooks such that the probability of error, the key leakage, and the uniformity of the keys tend to zero as $(n,m)\to\infty$.

The bounds on the cardinality of the alphabets $\cU$, $\cV$, $\cQ$, and $\cT$ follow from Fenchel--Eggleston--Carath{\'e}odory's theorem and the standard cardinality bounding technique~\cite[Appendix C]{gamal_network_2011}; therefore their proof is omitted.
This concludes the proof of Theorem~\ref{th:digital-key}.
\endIEEEproof

\section{Summary and Concluding Remarks}
\label{sec:summary}

In this work, we investigated the problem of secret key generation over a noisy channel in presence of correlated sources (independent of the main channel) at all terminals.
We introduced a novel coding scheme using separate source and channel components --which shares common roots with our previous works~\cite{villard_secure_2014, bassi_wiretap_2015}. 
With the use of two description layers on the source observed at the encoder, this scheme improves upon the existing works in the literature which only rely on one layer of description.

The corresponding achievable secret key rate was shown to be optimal for all classes of less-noisy sources and channels (Propositions~\ref{prop:sp_ln_z},~\ref{prop:sp_ln_e}, and~\ref{prop:sp_ln_yb}).
In Section~\ref{sec:bec_bsc_channel}, we compared the performance of the proposed scheme with a previously reported result for a simple binary model. Numerical computation of the corresponding bounds provided interesting insights on the regimes where the novel scheme outperforms the previous one.

This work, however, does not address the scenario where the sources and the noisy channel are correlated. The extension of the above mentioned result of Prabhakaran \emph{et al.}~\cite{prabhakaran_secrecy_2012} by using  two description layers is a natural consequence. Indeed, this extension --posterior to the short version of the present work in~\cite{bassi_isit_2016}-- has been recently addressed in~\cite{bunin_key_2018}. Using two description layers as introduced here, the proposed achievable scheme recovers the present inner bound for $\eta=1$ provided that the sources are independent of the channel.


\appendices
\section{Proof of Theorem~\ref{th:outer} (Outer Bound)}
\label{sec:Proof-OB}

The outer bound is derived by following similar steps to those in~\cite[Thm.~4]{bassi_wiretap_2015}, which assumed $\eta=1$. It is reproduced here for completeness.

Let $(\eta, R_k)$ be an achievable tuple according to Definition~\ref{def:achievability}, and $\epsilon>0$.
Then, there exists a $(2^{nR_k},n,m)$ secret key code $\mathsf{c}_n$ with functions $\varphi(\cdot)$, $\psi_a(\cdot)$, and $\psi_b(\cdot)$ such that
\begin{subequations}\label{eq:OB_functions}
\begin{align}
X^m     &= \varphi(A^n, R_r)\,, \label{eq:OB_functions1} \\
K       &= \psi_a(A^n, R_r)\,,  \label{eq:OB_functions2} \\
\hat{K} &= \psi_b(B^n, Y^m)\,,
\end{align}
\end{subequations}
that verify
\begin{subequations}
\begin{align}
\frac{m}{n}           &\leq \eta +\epsilon\,, \\
\PR{ K \neq \hat{K} } &\leq \epsilon\,, \label{eq:OB_cond1} \\
\I{K}{E^n Z^m}        &\leq n\epsilon\,, \label{eq:OB_cond2}\\
nR_k -\H{K}           &\leq n\epsilon\,, \label{eq:OB_cond3}
\end{align}
\end{subequations}
where we have dropped the conditioning on the codebook $\mathsf{c}_n$ from~\eqref{eq:OB_cond1}--\eqref{eq:OB_cond3} and all subsequent calculations for clarity.
Before continuing, we present the following remark that is useful to establish Markov chains between the random variables.

\vspace{1mm}
\begin{remark}
From the fact that random variables $A_i$, $B_i$, $E_i$ are independent across time and the channel $X\mapsto(Y,Z)$ is memoryless, the joint distribution of $(K, A^n, B^n, E^n, \allowbreak X^m, Y^m, Z^m)$ can be written as follows. For each $i\in[1:n]$ and each $j\in[1:m]$, we have
\begin{align}
\MoveEqLeft[1]
p(k,a^n,b^n,e^n,x^m,y^m,z^m) \nonumber\\
	&=      p(a^{i-1},b^{i-1},e^{i-1})\,p(a_i,b_i,e_i)\,p(a_{i+1}^n,b_{i+1}^n,e_{i+1}^n) \nonumber\\
	&\quad\, p(k,x^m|a^n)\, p(y^{j-1},z^{j-1}|x^{j-1})\,p(y_j,z_j|x_j) \nonumber\\
	&\quad\, p(y_{j+1}^m,z_{j+1}^m|x_{j+1}^m)\,, \label{eq:OB_pmf_markov}
\end{align}
where $P_\varphi(x^m|a^n) = \sum\nolimits_{\forall \,k} p(k,x^m|a^n)$ and $P_{\psi_a}(k|a^n) = \sum\nolimits_{\forall \,x^m} p(k,x^m|a^n)$.
\end{remark}
\vspace{1mm}

We may now carry on with the derivation of the outer bound.
First consider,
\begin{subequations}\label{eq:OB_rate1}
\begin{align}
nR_k &\leq \H{K} +n\epsilon \label{eq:OB_rate1a} \\
 &= \HC{K}{E^n Y^m} +\I{K}{E^n Y^m} +n\epsilon\\
 &\leq \HC{K}{E^n Y^m} +\I{K}{E^n Y^m} \nonumber\\
 &\quad-\I{K}{E^n Z^m} +2n\epsilon \label{eq:OB_rate1b} \\
 &= \HC{K}{E^n Y^m} +\IC{K}{Y^m}{E^n} \nonumber\\
 &\quad-\IC{K}{Z^m}{E^n} +2n\epsilon \displaybreak[2]\\
 &\leq \HC{K}{E^n Y^m} -\HC{K}{B^n Y^m} \nonumber\\
 &\quad +\IC{K}{Y^m}{E^n} -\IC{K}{Z^m}{E^n} +3n\epsilon \label{eq:OB_rate1c} \\
%
 &= {\underbrace{\IC{K}{B^n}{Y^m} -\IC{K}{E^n}{Y^m}}_{R_s}} \nonumber\\
 &\quad +{\underbrace{\IC{K}{Y^m}{E^n} -\IC{K}{Z^m}{E^n}}_{R_c}} +3n\epsilon\,, \label{eq:OB_rate1d}
\end{align}
\end{subequations}
where
\begin{itemize}
 \item \eqref{eq:OB_rate1a} stems from the uniformity of the keys~\eqref{eq:OB_cond3};
 \item \eqref{eq:OB_rate1b} is due to the security condition~\eqref{eq:OB_cond2}; and,
 \item \eqref{eq:OB_rate1c} follows from~\eqref{eq:OB_functions}, \eqref{eq:OB_cond1}, and Fano's inequality, $\HC{K}{B^n Y^m} \leq n\epsilon$.
\end{itemize}
We now study separately the ``source'' term $R_s$ and the ``channel'' term $R_c$. Hence,
\begin{subequations}\label{eq:OB_rate2}
\begin{align}
R_s &= \sum\nolimits_{i=1}^n \IC{K}{B_i}{Y^m B^{i-1}} -\IC{K}{E_i}{Y^m E_{i+1}^n} \nonumber\\
 &= \sum\nolimits_{i=1}^n \IC{K}{B_i}{Y^m B^{i-1} E_{i+1}^n} \nonumber\\
 &\quad -\IC{K}{E_i}{Y^m B^{i-1} E_{i+1}^n} \label{eq:OB_rate2a} \\
 &= \sum\nolimits_{i=1}^n \IC{V_i}{B_i}{U_i} -\IC{V_i}{E_i}{U_i} \label{eq:OB_rate2b} \displaybreak[2]\\
 &= n [ \IC{V_J}{B_J}{U_J J} -\IC{V_J}{E_J}{U_J J} ] \label{eq:OB_rate2c} \\
 &= n [ \IC{V}{B}{U} -\IC{V}{E}{U} ]\,, \label{eq:OB_rate2d}
\end{align}
\end{subequations}
where
\begin{itemize}
 \item \eqref{eq:OB_rate2a} is due to Csisz\'ar sum identity; 
 \item \eqref{eq:OB_rate2b} follows from the definition of the auxiliary RVs $U_i = (Y^m B^{i-1} E_{i+1}^n)$ and $V_i = (K U_i)$;
 \item \eqref{eq:OB_rate2c} introduces the auxiliary RV $J$ uniformly distributed over $[1:n]$ and independent of all the other variables; and,
 \item \eqref{eq:OB_rate2d} stems from the definition of random variables $U = (U_J J)$, $V = (V_J J)$, $B = B_J$, and $E = E_J$.
\end{itemize}
This establishes the ``source'' term in~\eqref{eq:OB_rate1d} with auxiliary RVs $(U,V)$ that satisfy the following Markov chain
\begin{equation}
 U_i \mkv V_i \mkv A_i \mkv (B_i E_i)\,. \label{eq:OB_mkv1}
\end{equation}
The first part of~\eqref{eq:OB_mkv1} is trivial given the definition $V_i = (K U_i)$, whereas the second part follows from the i.i.d. nature of the sources and that they are correlated to the main channel only through the encoder's input~\eqref{eq:OB_functions1}, see~\eqref{eq:OB_pmf_markov},
\begin{equation}
(K Y^m B^{i-1} E_{i+1}^n) \mkv A_i \mkv (B_i E_i)\,. 
\end{equation}

%
The ``channel'' term $R_c$ can be single-letterized similarly,
\begin{equation}
R_c = m [ \IC{T}{Y}{Q} -\IC{T}{Z}{Q} ]\,, \label{eq:OB_rate3}
\end{equation}
where we first define the auxiliary RVs $Q_i = (E^n Y^{i-1} Z_{i+1}^m)$ and $T_i = (K Q_i)$, we then introduce the auxiliary RV $L$ uniformly distributed over $[1:m]$, and we finally define $Q = (Q_L L)$, $T = (T_L L)$, $Y = Y_L$, and $Z = Z_L$.
The auxiliary RVs in this term, i.e., $(Q,T)$, satisfy the following Markov chain
\begin{equation}
 Q_i \mkv T_i \mkv X_i \mkv (Y_i Z_i)\,, 
\end{equation}
where the nontrivial part is due to the memoryless property of the channel and~\eqref{eq:OB_functions2}, provided the joint probability distribution satisfies~\eqref{eq:OB_pmf_markov}.
Since neither $Q$ nor $T$ appear on other parts of the outer bound, we may expand $R_c$ as
\begin{subequations}\label{eq:OB_rate3bis}
\begin{align}
 R_c &= m \sum_{q\in\cQ} p_Q(q) \left[ \IC{T}{Y}{Q=q} -\IC{T}{Z}{Q=q} \right] \\
 &\leq m \max_{q\in\cQ} \left[ \IC{T}{Y}{Q=q} -\IC{T}{Z}{Q=q} \right] \\
 &= m [ \I{T^\star}{Y} -\I{T^\star}{Z} ]\,,
\end{align}
\end{subequations}
where in the last step we choose auxiliary RV $T^\star \sim p_{T\vert Q}(\cdot\vert q)$.

Gathering~\eqref{eq:OB_rate1}, \eqref{eq:OB_rate2}, \eqref{eq:OB_rate3}, and~\eqref{eq:OB_rate3bis}, the rate of the secret key writes
\begin{equation}
R_k \leq \IC{V}{B}{U} -\IC{V}{E}{U} + \frac{m}{n} \big[ \I{T}{Y} -\I{T}{Z} \big] +3\epsilon\,.
\end{equation}
If we let $(n,m)\rightarrow\infty$ and take arbitrarily small $\epsilon$, we obtain the bound~\eqref{eq:outer-rate}.

In order to obtain~\eqref{eq:outer-cond}, we use the following Markov chain that is a consequence of~\eqref{eq:OB_functions1}, provided the joint probability satisfies~\eqref{eq:OB_pmf_markov}:
\begin{equation}
 (B^n E^n) \mkv A^n \mkv X^m \mkv (Y^m Z^m)\,. \label{eq:OB_mkv4}
\end{equation}
Due to the data processing inequality, we have
\begin{equation}
\I{A^n}{Y^m} \leq \I{X^m}{Y^m} \leq m\, \I{X}{Y}\,, \label{eq:OB_rate4bis}
\end{equation}
where in the last inequality we use the memoryless property of the channel.
Next, consider
\begin{subequations}\label{eq:OB_rate4}
\begin{align}
\I{A^n}{Y^m} &= \I{A^n B^n}{Y^m} \label{eq:OB_rate4a} \\
 &\geq \IC{A^n}{Y^m}{B^n} \\
 &= \IC{A^n}{K Y^m}{B^n} -\IC{A^n}{K}{B^n Y^m} \\
 &\geq \IC{A^n}{K Y^m}{B^n} -n\epsilon \label{eq:OB_rate4b} \\
 &\geq n [\IC{A}{V}{B} -\epsilon]\,, \label{eq:OB_rate4c}
\end{align}
\end{subequations}
where
\begin{itemize}
 \item \eqref{eq:OB_rate4a} follows from the Markov chain~\eqref{eq:OB_mkv4}; and,
 \item \eqref{eq:OB_rate4b} stems from $\HC{K}{B^n Y^m}\leq n\epsilon$ due to~\eqref{eq:OB_functions} and~\eqref{eq:OB_cond1}, and $\HC{K}{A^n B^n Y^m}\geq 0$.
\end{itemize}
For the last step, i.e., \eqref{eq:OB_rate4c}, consider
\begin{subequations}\label{eq:OB_rate5}
\begin{align}
\MoveEqLeft[1]
\IC{K Y^m}{A^n}{B^n} \nonumber\\
 &= \IC{K Y^m}{A^n E^n}{B^n} \label{eq:OB_rate5a} \\
 &= \sum\nolimits_{i=1}^n \IC{K Y^m}{A_i E_i}{B^n A_{i+1}^n E_{i+1}^n} \\
 &\geq \sum\nolimits_{i=1}^n \IC{K Y^m B^{i-1} E_{i+1}^n}{A_i E_i}{B_i} \label{eq:OB_rate5b} \displaybreak[2]\\
 &= \sum\nolimits_{i=1}^n \IC{V_i}{A_i E_i}{B_i} \label{eq:OB_rate5c} \displaybreak[2] \\
 &\geq \sum\nolimits_{i=1}^n \IC{V_i}{A_i}{B_i} \displaybreak[2] \\
 &= n\,\IC{V_J}{A_J}{B_J J} \label{eq:OB_rate5d} \displaybreak[2]\\
 &= n\,\IC{V_J J}{A_J}{B_J} \label{eq:OB_rate5e} \\
 &= n\,\IC{V}{A}{B}\,, \label{eq:OB_rate5f}
\end{align}
\end{subequations}
where
\begin{itemize}
 \item \eqref{eq:OB_rate5a} stems from the Markov chain $(B^n E^n) \mkv A^n \mkv (K Y^m)$;
 \item \eqref{eq:OB_rate5b} follows from the sources being i.i.d., i.e., $(A_i E_i) \mkv B_i \mkv (B^{i-1} B_{i+1}^n A_{i+1}^n E_{i+1}^n)$;
 \item \eqref{eq:OB_rate5c} is due to the auxiliary RV $V_i = (K Y^m B^{i-1} E_{i+1}^n)$;
 \item \eqref{eq:OB_rate5d} introduces the auxiliary RV $J$ uniformly distributed over $[1:n]$ and independent of all the other variables;
 \item \eqref{eq:OB_rate5e} follows from the independence of $J$ and $(A_J B_J)$; and,
 \item \eqref{eq:OB_rate5f} stems from the definition of random variables $V = (V_J J)$, $B = B_J$, and $A = A_J$.
\end{itemize}

Putting~\eqref{eq:OB_rate4bis} and~\eqref{eq:OB_rate4} together, we obtain:
\begin{equation}
\IC{V}{A}{B} \leq \frac{m}{n} \I{X}{Y} +\epsilon\,,
\end{equation}
which gives the condition~\eqref{eq:outer-cond} as we let $(n,m)\rightarrow\infty$ and take an arbitrarily small $\epsilon$.

Although the definition of the auxiliary RVs $(TUV)$ used in the proof makes them arbitrarily correlated, the bounds~\eqref{eq:outer-rate} and~\eqref{eq:outer-cond} only depend on the \emph{marginal} PDs $p(tx)$ and $p(uv\vert a)$. Consequently, we can restrict the set of possible joint PDs to~\eqref{eq:outer-pmf}, i.e., independent source and channel variables, and still achieve the maximum.

The bound on the cardinality of the alphabets $\cT$, $\cU$, and $\cV$ follow from Fenchel--Eggleston--Carath{\'e}odory's theorem and the standard cardinality bounding technique~\cite[Appendix C]{gamal_network_2011}; therefore their proof is omitted.
This concludes the proof of Theorem~\ref{th:outer}.
\endIEEEproof

\section{Proof of Proposition~\ref{prop:bec_bsc_sep_1l}}
\label{sec:Proof-BEC_BSC_1L}

For completeness, we first present the inner bound from~\cite[Thm.~4]{khisti_secret-key_2012} but rewritten using the notation of the present work:
\begin{subequations}\label{eq:inner_1l}
\begin{gather}
 R_k \leq \max_{p(x)p(v|a)} \big\{ \I{V}{B} -\I{V}{E} +\eta\, \IC{X}{Y}{Z} \big\} \label{eq:inner_1l-a}\\
 \textnormal{subject to }\ \IC{V}{A}{B} \leq \eta\,\I{X}{Y}\,. \label{eq:inner_1l-b}
\end{gather}
\end{subequations}
In the sequel, we assume $\eta=1$.

The main channel in the system model depicted in Fig.~\ref{fig:bec_bsc_channel_a} is not only degraded but also $Y$ equals $X$; thus, the last term on the r.h.s. of~\eqref{eq:inner_1l-a} may be expanded as follows
\begin{equation}
 \IC{X}{Y}{Z} = \HC{X}{Z} = \H{X} +\HC{Z}{X} -\H{Z}\,.
\end{equation}
Since $X$ is the input of a BSC of parameter $\zeta$ and output $Z$, it is clear that
\begin{equation}
 \IC{X}{Y}{Z} \leq \HC{Z}{X} = h_2(\zeta)\,, \label{eq:proof-bec_bsc_1l-1}
\end{equation}
with equality if and only if $X\sim\cB\left(\frac12\right)$. Moreover, this choice of $X$ maximizes the r.h.s. of~\eqref{eq:inner_1l-b} and makes the condition redundant:
\begin{equation}
 \IC{V}{A}{B} \leq \HC{A}{B} = \beta \H{A} = \beta \leq 1 = \H{X} \,,
\end{equation}
given that $A\sim\cB\!\left(\tfrac12\right)$ and $0\leq \beta\leq 1$.

It remains to be determined what the maximizing value of the first two terms on the r.h.s. of~\eqref{eq:inner_1l-a} is. Let us first assume that $B$ is \emph{more capable} than $E$, i.e., $0\leq \beta< h_2(\epsilon)$ according to Remark~\ref{rk:bec_bsc_channel_prop}. Then, we may write
\begin{subequations}\label{eq:proof-bec_bsc_1l-2}
\begin{align}
\MoveEqLeft[1]
\I{V}{B} -\I{V}{E} \nonumber\\
 &= \I{A}{B} -\I{A}{E} - \big[ \IC{A}{B}{V} -\IC{A}{E}{V} \big] \nonumber\\
 &\leq \I{A}{B} -\I{A}{E} \\
 &= \HC{A}{E} -\HC{A}{B} \\
 &= h_2(\epsilon) -\beta\,,
\end{align}
\end{subequations}
where the inequality is due to $\IC{A}{B}{V} \geq \IC{A}{E}{V}$ for all $p(v,a)$ given the more capable assumption. The bound~\eqref{eq:proof-bec_bsc_1l-2} holds with equality if and only if $V=A$. 
We also note that~\eqref{eq:proof-bec_bsc_1l-2} is a monotonically decreasing function of $\beta$ and it is zero when $\beta= h_2(\epsilon)$. 
For $\beta>h_2(\epsilon)$, the bound~\eqref{eq:proof-bec_bsc_1l-2} is no longer valid; however, we can rightfully argue that as Bob's source degrades while Eve's remains the same, it is not possible to obtain more secret bits from the sources than for $\beta=h_2(\epsilon)$. Therefore, for $\beta>h_2(\epsilon)$,
\begin{equation}
 \I{V}{B} -\I{V}{E} \leq 0\,, \label{eq:proof-bec_bsc_1l-3}
\end{equation}
which holds with equality if and only if $V=\emptyset$.

Combining~\eqref{eq:inner_1l}, \eqref{eq:proof-bec_bsc_1l-1}, \eqref{eq:proof-bec_bsc_1l-2}, and~\eqref{eq:proof-bec_bsc_1l-3}, we obtain
the bound in~\eqref{eq:bec_bsc_inner_sep_1l}.
This concludes the proof of Proposition~\ref{prop:bec_bsc_sep_1l}.
\endIEEEproof

\section{Proof of Lemma~\ref{lem:lem_bnd_ps}}
\label{sec:Proof-Lemma-Bound_ps}

According to the encoding procedure detailed in Section~\ref{ssec:key_encoding}, the index $S$ is chosen uniformly among all the jointly typical codewords or, if there is no jointly typical codeword, uniformly on the whole codebook. We may thus characterize $p_{S_c}(1)$ as
\begin{equation}
 p_{S_c}(1) = \sum_{a^n \in \typ{n}{A}} \frac{p(a^n)}{\PR{\typ{n}{A}}}\ \Upsilon_{a^n}\,, 
 \label{eq:lemma-bound_11}
\end{equation}
where
\begin{equation}
 \Upsilon_{a^n} = \frac{\nu_1}{1+\sum_{i=2}^{|\cS|} \nu_i} +|\cS|^{-1} \prod_{i=1}^{|\cS|} (1 -\nu_i)\,,
 \label{eq:lemma-bound_12}
\end{equation}
and $\nu_i$ is the event that the codeword $v^n(i)$ is jointly typical with the source sequence $a^n$, i.e.,
\begin{multline}
 \nu_i \triangleq \mathds{1} \big\{ v^n(i) \in \typ{n}{V\vert u^n, a^n} \\ \cond v^n(i) \in \typ{n}{V\vert u^n}, u^n \in \typ{n}{U\vert a^n} \big\}\,.
\end{multline}
The first term in~\eqref{eq:lemma-bound_12} distributes the probability of each sequence $a^n \in \typ{n}{A}$ uniformly among all the jointly typical codewords, while the second term in~\eqref{eq:lemma-bound_12} distributes this probability uniformly among all codewords in \cS, given that no one was jointly typical with $a^n$.
It is not hard to see that the expected value of $\nu_i$ is
\begin{equation}
 \bE_{\cC}[\nu_i] 
 =  \frac{| \typ{n}{V\vert u^n, a^n} |}{| \typ{n}{V\vert u^n} |} 
 \triangleq \gamma\,,
\end{equation}
for some $(u^n, a^n)\in\typ{n}{UA}$.

The expected value of~\eqref{eq:lemma-bound_11} depends on the behavior of $\Upsilon_{a^n}$. Each $\nu_i$ is a Bernoulli RV with $\bE_{\cC}[\nu_i]=\gamma$ and it is independent of the other $\nu_i$'s. Let us define
\begin{equation}
 \nu = \sum\nolimits_{i=2}^{|\cS|} \nu_i\,,
\end{equation}
then $\nu$ is a Binomial RV, and thus, for $j\in[0:|\cS|-1]$,
\begin{equation}
 p_\nu(j) = \binom{|\cS|-1}{j} \gamma^j (1-\gamma)^{|\cS|-1-j}\,.
\end{equation}
After some manipulations, it is possible to show that
\begin{equation}
 \bE_{\cC}\!\left[ \!\frac{1}{1+\nu} \right] = \frac{1-(1-\gamma)^{|\cS|}}{\gamma\, |\cS|}\,.
\end{equation}
Hence,
\begin{equation}
 \bE_{\cC}[\Upsilon_{a^n}]
 = \bE_{\cC}\!\left[ \frac{\nu_1}{1+ \nu} +\frac{1}{|\cS|} \prod_{i=1}^{|\cS|} (1 -\nu_i) \right]
 = \frac{1}{|\cS|}\,,
\end{equation}
and consequently, the expected value of~\eqref{eq:lemma-bound_11} is
\begin{equation}
 \bE_{\cC}[p_{S_c}(1)]
 = \bE_{\cC}[\Upsilon_{a^n}]
 = |\cS|^{-1}\,.
\end{equation}

Noting that $\Upsilon_{a^n}$ and $\Upsilon_{{a^n}'}$ are independent variables given different sequences $a^n$ and ${a^n}'$, and that $(\Upsilon_{a^n})^2\leq \Upsilon_{a^n}$, we obtain
\begin{equation}
 \bE_{\cC}[(p_{S_c}(1))^2]
 \leq 2^{-n[\H{A}-\xi]} |\cS|^{-1} + |\cS|^{-2}\,,
\end{equation}
for some $\xi>0$. Therefore,
\begin{equation}
 \textnormal{Var}[p_{S_c}(1)] \leq 2^{-n[\H{A}-\xi]} |\cS|^{-1}\,,
\end{equation}
and in view of Chebyshev's inequality,
\begin{align*}
\PR{ \big| p_{S_c}(1) -|\cS|^{-1} \big| \geq \varepsilon_1\, |\cS|^{-1} } 
 &\leq \varepsilon_1^{-2} 2^{-n[\H{A} -\xi]} |\cS| \\
 &= \varepsilon_1^{-2} 2^{-n[\H{A} -S_2 -\xi]}\,.
\end{align*}
This probability converges exponentially fast towards zero if $S_2<\H{A} -\xi$.
This concludes the proof of Lemma~\ref{lem:lem_bnd_ps}.
\endIEEEproof

\section{Proof of Lemma~\ref{lem:lem_fano_eve}}
\label{sec:Proof-Lemma-Fano_Eve}

Let us modify the problem definition and then extend the scheme of Theorem~\ref{th:digital-key} by introducing two virtual users who observe the source sequence $E^n$. The first user has access to the index $r_1$ as side information and we require that it decodes the codeword $U^n$. On the other hand, the second user has access to a different side information (which contains $U^n$) and we require that it decodes the codeword $V^n$. The keen reader can immediately see that we may bound the entropies in the statement of the lemma using Fano's inequality if the probability of error at the virtual users tend to zero.

Before proceeding, we note that the entropy in~\eqref{eq:lem_fano_eve2} has $(Z^m, K_2)$ in the conditioning. These variables are related to the \emph{channel} and they affect the entropy of the \emph{source}-related variable $V^n$ through the knowledge they provide about the index $r_p$. In the sequel, we first characterize the decrease on the entropy of $r_p$ and we then proceed to analyze the probability of error of the virtual users.

Let us introduce the random variable $\Upsilon$, such that
\begin{equation}
\Upsilon \triangleq \ind{(Q^m, Z^m) \in \typ{m}{QZ}}\,.
\end{equation}
Then, using the binary variable $\Upsilon$, it follows that
\begin{align}
\MoveEqLeft[1]
\HC{V^n}{\cC E^n Z^m U^n r_2 K_1 K_2} & \nonumber \\
 &\leq 1 + \HC{V^n}{\cC E^n Z^m U^n r_2 K_1 K_2 \Upsilon} \nonumber\\
 &\leq 1 + \HC{V^n}{\cC E^n Z^m U^n r_2 K_1 K_2, \Upsilon=1} + nS_2 \delta\,, \label{eq:lemma-fano-1b}
\end{align}
where the last inequality is due to $\HC{V^n}{\cC U^n}\leq nS_2$ and $\PR{\Upsilon=0}\leq\delta$.

In order to bound~\eqref{eq:lemma-fano-1b}, we observe that, although $r_p\in[1:2^{nR_p}]$, the index has only a non-zero probability in a smaller subset of indices given the conditioning on $Z^m$, $r_1$ (known through $U^n$ and $\cC$), $r_2$, $K_2$, and $\Upsilon=1$.
For a specific codebook $\mathsf{c}_n$ (which defines the codewords $q^m(\cdot)$ and $t^m(\cdot)$), a channel output $z^m$, and the indices $r_1$, $r_2$, and $k_2$, let us define the set of possible indices $r_p$ as
\begin{multline}
\cS_R \triangleq\{ r_p : t^m(r_1, r_2, r_p, k_2, r_f) \in\typ{m}{T|q^m(r_1,r_2),z^m} \\ \textnormal{ for some } r_f  \} \,.
\label{eq:lemma-fano-sr}
\end{multline}
In principle, the size of this set depends on the particular codebook, channel output, and indices chosen. However, for sufficiently large $n$, the following lemma shows that the cardinality of $\cS_R$ is close to its mean value for almost all codebooks.

\vspace{1mm}
\begin{lemma}\label{lem:lem_bnd_SR}
Let $\varepsilon_1, \varepsilon_5, \varepsilon_6>0$, and let $\chi$ be a function of the codebook $\mathsf{c}_n$, the sequence $z^m$, and the indices $r_1$, $r_2$, and $k_2$ (not shown explicitly) defined as
\begin{multline}
 \chi(\mathsf{c}_n, z^m) \triangleq \\ \ind{ \big| S_R -[ R_p +R_f -(\eta+\epsilon)\IC{T}{Z}{Q}] \big| \geq \varepsilon_5 }\,, 
\end{multline}
where $S_R \triangleq \frac{1}{n}\log |\cS_R|$ and the set $\cS_R$ is defined in~\eqref{eq:lemma-fano-sr}.
Then, $\PR{\chi(\cC, Z^m)=1}\leq \varepsilon_6$ for sufficiently large $n$ if $R_p +R_f >(\eta+\epsilon)\IC{T}{Z}{Q} +\varepsilon_1$.
\end{lemma}
\begin{IEEEproof}
See Appendix~\ref{sec:Proof-Lemma-Bound_Sr}.
\end{IEEEproof}
\vspace{1mm}

We may thus write,
\begin{subequations}\label{eq:lemma-fano-2}
\begin{align}
\MoveEqLeft[1]
\HC{V^n}{\cC E^n Z^m U^n r_2 K_1 K_2, \Upsilon=1} & \nonumber \\
 &\leq \HC{V^n}{\cC E^n Z^m U^n r_2 K_1 K_2, \Upsilon=1, \chi=0} \nonumber\\
 &\quad + nS_2\varepsilon_6\,, \label{eq:lemma-fano-2a}\\
 &= \HC{V^n}{\cC E^n Z^m U^n r_2 K_1 K_2, \Upsilon=1, r_p\in\cS_R, \chi=0} \nonumber\\
 &\quad + nS_2\varepsilon_6\,, \label{eq:lemma-fano-2b}\\
 &\leq \HC{V^n}{\cC E^n U^n r_2 K_1, r_p\in\cS_R, \chi=0} + nS_2\varepsilon_6\,,\! \label{eq:lemma-fano-2c}
\end{align}
\end{subequations}
where
\begin{itemize}
 \item \eqref{eq:lemma-fano-2a} follows from $\HC{V^n}{\cC U^n}\leq nS_2$ and Lemma~\ref{lem:lem_bnd_SR}, where $\chi$ denotes $\chi(\cC, Z^m)$; and,
 \item \eqref{eq:lemma-fano-2b} is due to $r_p\in\cS_R$ being a function of $(\cC, Z^m, r_1, \allowbreak r_2, K_2, \Upsilon=1)$.
\end{itemize}
 
In light of~\eqref{eq:lemma-fano-2c}, we define the side information of the second virtual user as $(u^n(s_1), r_2, k_1, \allowbreak r_p\in\cS_R)$.
According to the random codebook generation procedure, the number of codewords $V^n(\cdot)$ in a particular sub-bin $\tilde{\cB}_2$ is $|\tilde{\cB}_2(s_1,r_2,r_p)| = 2^{n(S_2-R_2-R_p)}$; thus, conditioned on $(u^n(s_1), \allowbreak r_2, r_p\in\cS_R, \chi=0)$, there are at most
\begin{align*}
\sum\nolimits_{r_p\in\cS_R} |\tilde{\cB}_2(s_1,r_2,r_p)| &= 2^{\log|\cS_R|} 2^{n (S_2-R_2-R_p)} \nonumber\\
 &\leq 2^{n [S_2-R_2 +R_f -(\eta+\epsilon)\IC{T}{Z}{Q} + \varepsilon_5 ]} 
\end{align*}
distinct codewords $V^n(\cdot)$.
These codewords will be evenly distributed in the sub-bins $\bar{\cB}_2$, given the symmetry of the random codebook generation and the independence in the creation of the sub-bins $\tilde{\cB}_2$ and $\bar{\cB}_2$, if
\begin{equation}
 R_{k_1} < S_2-R_2 +R_f -(\eta+\epsilon)\IC{T}{Z}{Q} + \varepsilon_5\,.
\end{equation}
The reader may verify that this is true due to~\eqref{eq:th1-enc2} and~\eqref{eq:th1-keyleak_7}. Therefore, using the side information $(u^n(s_1), r_2, k_1, r_p\in\cS_R)$, the second virtual user can construct a set of possible codewords $V^n(\cdot)$ defined as
\begin{equation}
 \cS_V \triangleq \bigcup_{\smash{r_p\in\cS_R}} \tilde{\cB}_2(s_1,r_2,r_p) \cap \bar{\cB}_2(s_1,r_2,k_1)\,,
\end{equation}
where the number of codewords is at most
\begin{equation}
\left| \cS_V  \right| \leq 2^{n [S_2 -R_2 -R_{k_1} +R_f -(\eta+\epsilon)\IC{T}{Z}{Q} + \varepsilon_5 ]}\,.
\end{equation}

We are finally ready to state the modified problem definition. Let virtual user 1 decode the codeword $u^n(s_1)\in\cB_1(r_1)$ using the source sequence $e^n$, i.e., it looks for the unique index $s_1 \equiv \hat{s}_1$ such that $u^n(\hat{s}_1)\in\cB_1(r_1)$ and
\begin{equation}
 \left( u^n(\hat{s}_1), e^n \right) \in \typ{n}{UE}\,.
\end{equation}
The probability of error in decoding is arbitrarily small as $n\to\infty$ if 
\begin{equation}
 S_1 -R_1 < \I{U}{E} -\delta'\,. \label{eq:lemma-fano-3}
\end{equation}
On the other hand, let virtual user 2 decode the codeword $v^n(s_1,s_2)\in\cS_V$ using the source sequence $e^n$, i.e., it looks for the unique index $s_2 \equiv \hat{s}_2$ such that $v^n(s_1,\hat{s}_2)\in\cS_V$ and
\begin{equation}
 \left( v^n(s_1,\hat{s}_2), e^n \right) \in \typ{n}{VE|u^n(s_1)}\,.
\end{equation}
The probability of error in decoding is arbitrarily small as $n\to\infty$ if 
\begin{multline}
 S_2 -R_2 -R_{k_1} +R_f -(\eta+\epsilon) \IC{T}{Z}{Q} +\varepsilon_5 \\ < \IC{V}{E}{U} -\delta'\,. \label{eq:lemma-fano-4}
\end{multline}

To sum up, if~\eqref{eq:lemma-fano-3} and~\eqref{eq:lemma-fano-4} hold true, the probability of error in decoding at the virtual users is arbitrarily small as $n\to\infty$. Therefore, using Fano's inequality, we have
\begin{subequations}\label{eq:lemma-fano-5}
\begin{align}
 \HC{U^n}{\cC E^n r_1} &\leq n\epsilon_n\,, \\
 \HC{V^n}{\cC E^n U^n r_2 K_1, r_p\in\cS_R, \chi=0} &\leq n\epsilon_n\,,
\end{align}
\end{subequations}
where $\epsilon_n$ denotes a sequence such that $\epsilon_n\to 0$ as $n\to\infty$.
Joining~\eqref{eq:lemma-fano-1b}, \eqref{eq:lemma-fano-2}, and~\eqref{eq:lemma-fano-5}, we recover the statement of the lemma.
This concludes the proof of Lemma~\ref{lem:lem_fano_eve}.
\endIEEEproof

\subsection{Proof of Lemma~\ref{lem:lem_bnd_SR}}
\label{sec:Proof-Lemma-Bound_Sr}

Before analyzing the set of possible indices $r_p$, let us first concentrate on characterizing the set of possible codewords $t^m(\cdot)$.
Since the indices $r_1$, $r_2$, and $k_2$ are fixed, there are only $2^{n(R_p+R_f)}$ codewords to choose from.
Moreover, given a specific codebook $\mathsf{c}_n$ generated according to the procedure from Section~\ref{ssec:key_codebook}, the indices $r_1$ and $r_2$ fix the codeword $q^m(r_1,r_2)$; thus, we simplify the notation and the codebook is composed of $q^m\in\typ{m}{Q}$ and $t^m(r)\in\typ{m}{T|q^m}$, where $r\in[1:2^{n(R_p+R_f)}]$.
The set $\cS_T$ of possible codewords $t^m(r)$ is then defined as
\begin{equation}
\cS_T \triangleq\{ t^m(r) : t^m(r) \in\typ{m}{T|q^m,z^m} \} \,.
\end{equation}
Then, according to the random codebook generation,
\begin{align}
 \bE_{\cC Z^m}[|\cS_T|] &= \sum\limits_{r=1}^{2^{n(R_p+R_f)}} \bE_{\cC Z^m}[ \ind{ T^m(r) \in\typ{m}{T|q^m,z^m} } ] \nonumber\\
  &= 2^{n(R_p+R_f-\alpha)}\,,
\end{align}
where,
\begin{equation*}
 2^{-n\alpha} \triangleq \PR{ T^m(1) \in\typ{m}{T|q^m,z^m} } = \frac{|\typ{m}{T\vert q^m,z^m}|}{|\typ{m}{T\vert q^m}|}\,,
\end{equation*}
for some $(q^m,z^m)\in\typ{m}{QZ}$.
The value of $\alpha$ may be bounded using standard bounds for the cardinality of typical sets, yielding
\begin{equation}
 (\eta+\epsilon)\IC{T}{Z}{Q} -\varepsilon_1 \leq \alpha \leq (\eta+\epsilon)\IC{T}{Z}{Q} +\varepsilon_1\,,
\end{equation}
for some $\varepsilon_1>0$.

Similarly, we may calculate
\begin{equation*}
 \bE_{\cC Z^m}[|\cS_T|^2] = 2^{2n(R_p+R_f-\alpha)} +2^{n(R_p+R_f-\alpha)} (1- 2^{-n\alpha}  )\,,
\end{equation*}
and finally,
\begin{equation}
 \textnormal{Var}[|\cS_T|] \leq 2^{n(R_p+R_f-\alpha)}\,.
\end{equation}

We may now use Chebyshev's inequality to bound the value of $|\cS_T|$,
\begin{multline}
 \PR{ \big| |\cS_T| -\bE_{\cC Z^m} [|\cS_T|] \big| \geq \varepsilon_2\, \bE_{\cC Z^m} [|\cS_T|] } \\ \leq \varepsilon_2^{-2} 2^{-n(R_p +R_f -\alpha)}\,, \label{eq:lemma-bound-sr_chev}
\end{multline}
for some $\varepsilon_2>0$.
This probability tends to zero exponentially fast with $n$ if $R_p +R_f > (\eta+\epsilon)\IC{T}{Z}{Q} +\varepsilon_1$.
Taking the logarithm in the argument of the probability of~\eqref{eq:lemma-bound-sr_chev} we obtain
\begin{equation}
 \PR{ \bigg| \frac{1}{n}\log|\cS_T| -\beta \bigg| \geq \varepsilon_3 } \leq \varepsilon_4\,,
 \label{eq:lemma-bound-sr_chev2}
\end{equation}
for some $\varepsilon_3 \geq \varepsilon_1 +\frac{1}{n}\log(1 +\varepsilon_2)$ and $\varepsilon_4 \geq \varepsilon_2^{-2} 2^{-n(\beta -\varepsilon_1)}$, where
\begin{equation}
 \beta \triangleq R_p +R_f -(\eta+\epsilon)\IC{T}{Z}{Q}\,.
\end{equation}

We note that~\eqref{eq:lemma-bound-sr_chev2} provides an estimate on the second virtual receiver's uncertainty on the actual transmitted codeword $T^m(\cdot)$, i.e., the set $\cS_T$, rather than the index $r_p$, i.e., the set $\cS_R$. 
In order to bound the latter, consider the following
\begin{subequations}\label{eq:lemma-bound-sr_3}
\begin{align}
\MoveEqLeft[1]
 \bE_{\cC Z^m} [\log|\cS_R|] \nonumber\\
 &\leq \bE_{\cC Z^m}\big[\log|\cS_R|\, \big| |\cS_R|\leq 2^{n(\beta +\varepsilon_3)}\big] +nR_p \varepsilon_4 \label{eq:lemma-bound-sr_3a}\\
  &\leq n(\beta-\varepsilon_5) p_s + n(\beta+\varepsilon_3) (1-p_s) +nR_p \varepsilon_4 \label{eq:lemma-bound-sr_3b}\\
  &= n[\beta +\varepsilon_3 +\varepsilon_4 R_p -p_s(\varepsilon_3 +\varepsilon_5)]\,, 
\end{align}
\end{subequations}
where
\begin{itemize}
 \item \eqref{eq:lemma-bound-sr_3a} follows from having at most $2^{nR_p}$ indices $r_p$, the fact that $|\cS_R|\leq|\cS_T|$ (e.g. some indices might be repeated), and~\eqref{eq:lemma-bound-sr_chev2}; and,
 \item \eqref{eq:lemma-bound-sr_3b} is due to the definition $p_s\triangleq\PR{ |\cS_R|\leq 2^{n(\beta-\varepsilon_5)} }$, for some $\varepsilon_5>0$.
\end{itemize}
On the other hand, consider the following lower bound derived from~\eqref{eq:th1-keyleak_4}:
\begin{equation}
 \bE_{\cC Z^m} [\log|\cS_R|] \geq \HC{r_p}{\cC Z^m r_1 r_2 K_2} \geq n (\beta -\varepsilon')\,, \label{eq:lemma-bound-sr_4}
\end{equation}
where the first inequality is due to the definition of the set $\cS_R$ in~\eqref{eq:lemma-fano-sr} and the fact that the uniform distribution maximizes the entropy.
Joining~\eqref{eq:lemma-bound-sr_3} and~\eqref{eq:lemma-bound-sr_4} we obtain,
\begin{equation}
 p_s \leq \frac{\varepsilon' +\varepsilon_3 +\varepsilon_4 R_p}{\varepsilon_3 +\varepsilon_5} < \varepsilon_6\,, \label{eq:lemma-bound-sr_5}
\end{equation}
where the last inequality holds if $\varepsilon_5 \gg \max\{\varepsilon', \varepsilon_3, \varepsilon_4 R_p\}$. For a sufficiently large $n$, it is always possible to find such a $\varepsilon_5$.
Finally, the lemma's statement is recovered using~\eqref{eq:lemma-bound-sr_chev2} (jointly with the fact that $|\cS_R|\leq|\cS_T|$) and~\eqref{eq:lemma-bound-sr_5}.
This concludes the proof of Lemma~\ref{lem:lem_bnd_SR}.
\endIEEEproof

%

\bibliographystyle{IEEEtran}
\bibliography{IEEEabrv,biblio}

\end{document}